\documentclass[aps,prb,twocolumn,superscriptaddress,showpacs,floatfix]{revtex4}

\usepackage{amsmath,amssymb}
\usepackage{graphicx}

\begin{document}

\title{Electronic topological transition in sliding bilayer graphene}

\author{Young-Woo Son}
\email{hand@kias.re.kr}
\affiliation
{Korea Institute for Advanced Study, Seoul 130-722, Korea.}
\author{Seon-Myeong Choi}
\affiliation
{Korea Institute for Advanced Study, Seoul 130-722, Korea.}
\affiliation
{Department of Physics,
  Pohang University of Science and Technology, Pohang 790-784, Korea.}
\author{Yoon Pyo Hong}
\author{Sungjong Woo}
\affiliation
{Korea Institute for Advanced Study, Seoul 130-722, Korea.}
\author{Seung-Hoon Jhi}
\email{jhish@postech.ac.kr}
\affiliation
{Department of Physics,
  Pohang University of Science and Technology, Pohang 790-784, Korea.}
\affiliation
{Division of Advanced Materials Science,
  Pohang University of Science and Technology, Pohang 790-784, Korea.}

\begin{abstract}
We demonstrate theoretically that the topology of energy bands and Fermi surface in bilayer graphene 
undergoes a very sensitive transition when an extremely tiny lateral interlayer 
shift occurs in arbitrary directions. 
The phenomenon originates from a generation of an effective non-Abelian 
vector potential in the Dirac Hamiltonian by the sliding motions. 
The characteristics of the transition such as pair annihilations of massless Dirac fermions 
are dictated by the sliding direction owing to a unique interplay between the effective non-Abelian 
gauge fields and Berry's phases associated with massless electrons. The transition 
manifests itself in various measurable quantities such as anomalous density of states, 
minimal conductivity, and distinct Landau level spectrum.
\end{abstract}
\pacs{73.22.Pr,71.20.-b,81.05.ue,61.48.Gh}


\maketitle
\section{Introduction}

Changes in the topology of Fermi surfaces known as Lifshitz transition~\cite{lifshitz} alter 
physical properties of metals signifcantly~\cite{lifshitz,blanter,varlamov}. 
Though such an electronic topological 
transition (ETT) has been pursued in various materials, its realization
requires large external perturbations such as alloying or applying high pressure 
that hinder clear detections of the transition~\cite{lifshitz,blanter,varlamov,armitage}.  
Recent progress in measuring low energy electronic structures of bilayer 
graphene (BLG)~\cite{novoselov,li,yacoby1,yacoby2}
provides a new opportunity to explore the ETT 
because of its unique electronic structures 
and because of the possible noninvasive control of chemical potential of the system~\cite{lemonik}.

In BLG, two coupled hexagonal lattices of carbon atoms are arranged 
according to Bernal stacking~\cite{mccann,ando,nilsson,cserti,min,neto}. 
Because BLG has a large degeneracy at the charge neutral point~\cite{mccann,ando,nilsson,cserti,min,neto},
there have been intense discussions on possible many-body effects 
in the system~\cite{yacoby2,lemonik,silva,levitov,polini,vafek}
Moreover, since electrons in a single layer graphene (SLG) 
behave as relativistic massless fermions~\cite{neto},
BLG provides a unique playground to control interactions between 
relativistic particles coupled with the relative mechanical motions of two layers.
Hence, the effects of rotational stacking fault on physical properties of BLG
have been studied extensively~\cite{lopes,shallcross,li,hass}:
however, the effect of sliding one layer with respect to the other has not.
This mechanical motion is important because
the interactions between the two layers are sensitive to 
the deviation from Bernal stacking~\cite{lopes,shallcross,hass} 
and extremely small sliding will change its low energy electronic structures significantly. 

In this paper, we predict a very sensitive topological
change in the energy bands and Fermi surfaces of BLG when
sliding motion or interlayer shear occurs.  
It is demonstrated that a peculiar coupling 
between the effective gauge potential with SU(4) symmetry
generated by sliding motions and Berry's phase of massless Dirac fermions play
a crucial role to change the topology of low energy bands of BLG.
It gives rise to either pair annihilations of massless Dirac fermions
or generations of fermions by absorbing fermions with topological charges,
depending on sliding directions. This will offer new 
opportunities to realize the ETT driven by non-Abelian gauge 
fields with gentle maneuverable mechanical operations.

We start with a detailed description of our first-principles calculation methods
including a correction for interlayer dispersive forces. The energetics
and changes of the interlayer distance for sliding BLG are presented also. 
Then, the low energy electronic
structures obtained by the calculation are discussed when BLG experiences
very small sliding between the two layers. 
The next three sections introduce a model Hamiltonian for the system 
and discuss the role of gauge potential and Berry's phase for the changes in low energy
electronic structures. Several spectroscopic consequences will be discussed in the final section.

\section{First-principles calculation methods and atomic structures}

Our electronic structure calculation employs the first-principles
self-consistent pseudopotential method~\cite{espresso} using
the generalized gradient approximation (GGA) for exchange-correlation
functional emplemented by Perdew, Burke, and Ernzerhof (PBE)~\cite{pbe}.
A kinetic energy cutoff for wavefunction of 65 Rydberg is
employed and a plane-wave basis set is used.
The ion core of carbon atom is described by ultrasoft pseudopotential~\cite{vanderbilt}.
A $k$-point sampling of $60\times 60\times 1$
$k$ points uniformly distributed in the two-dimensional Brilouin zone (BZ)
is used in self-consistent calculations
and $150\times 150\times 1$ $k$ points is sampled to obtain electronic
energy bands on a rectangular grid
of 1.2\% of the first BZ size around the $K$-point.
Since we have dealt with sub-Angstrom displacements of atoms,
we have tested our calculation as increasing
the cutoff to 120 Rydberg (corresponding to a kinetic energy cutoff of
480 Rydberg for charge density and potential)
finding no differences in results.
The total charge was calculated by using the Marzari-Vanderbilt
cold smearing scheme~\cite{marzari}.
All atomic coordinates are relaxed and
the nearest-neighbor carbon-carbon distance ($a_c$) in a single
layer of graphene (SLG) is found to be 1.425~\AA.
We also perform first-principles calculations again for all different stacking
geometries by using another computational package~\cite{kresse} finding no difference.

Since the GGA cannot describe the interlayer interaction
between graphene properly, we have employed a seimemprical addition
of van der Waals (vdW) forces to our calculations following
Grimme's proposal~\cite{grimme}.
Within the semiemprical method, the total energy of the system
is $E_{\rm PBE+vdW}=E_{\rm PBE}+E_{\rm vdW}$ where $E_{\rm PBE}$
is the total energy from GGA functional of PBE
and the total energy given by dispersive forces is $E_{\rm vdW}$
which can be written as
$ E_{\rm vdW}=-\frac{1}{2}\sum_{i,j}C_{6ij}
\sum_{\vec R} f_{\rm damp}(|\vec{r}_{ij}+\vec{R}|)|\vec{r}_{ij}+\vec{R}|^{-6},
$
where $f_{\rm damp}
=s_6\cdot(1+e^{-d\cdot({|\vec{r}_{ij}+\vec{R}|}/{r_0}-1)})^{-1}$,
$\vec{r}_{ij}\equiv\vec{r}_i-\vec{r}_j$ is a vector for
carbon-carbon distance, $\vec R$ a lattice vector,
$s_6$ a scaling parameter, and $d$ a damping paramter, respectively~\cite{grimme,barone}.
The coefficient of dispersive forces of $C_{6ij}$
and a sum of vdW radii $r_0$ are computed for each pair
of atoms $i$ and $j$ such that
$C_{6ij}=\sqrt{C_{6i}C_{6j}}$ and $r_0=r_{0i}+r_{0j}$.
We have used $C_6=1.75$ J nm$^6$ mol$^{-1}$ and $r_0=1.451$~\AA~
for carbon atom suggested
by Grimme~\cite{grimme} and $d=20$, $s_6=0.65$ for our GGA calculations.
Here, $i$ and $j$ for $E_{\rm vdW}$ run through all atoms in the unit cell
and $\vec R$ satisfies a criteria of $|\vec{r}_{ij}+\vec{R}|<100$\AA.

\begin{figure}[t]
\includegraphics[width=1.0\columnwidth]{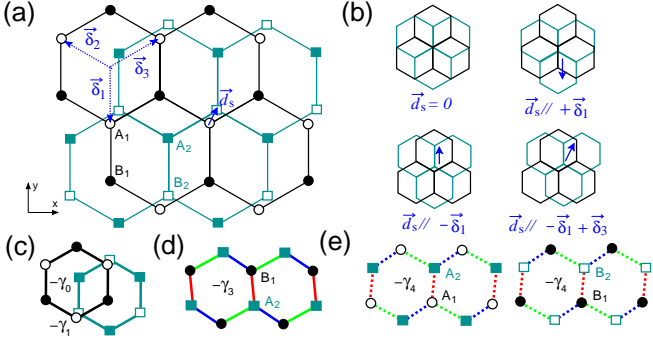}
\caption{(color online)
(a) In the top (black) and bottom layer (green), 
two sublattices are denoted by empty ($A_1$) 
and filled ($B_1$) circles and by empty ($B_2$) and filled ($A_2$) rectangles, respectively.
The bottom layer (layer 2) slides with repect to the top (layer 1) by $\vec d_s$. 
In the top, $\vec\delta_1=(0,-1)a_c$, $\vec\delta_2=(-\frac{\sqrt{3}}{2},\frac{1}{2})a_c$
and $\vec\delta_3=(\frac{\sqrt{3}}{2},\frac{1}{2})a_c$.
(b) Schematic diagrams for stacking under various sliding directions. 
(c) The nn intra- and inter-layer hopping parameter, $-\gamma_0$ and $-\gamma_1$. 
(d) The nnn inter-layer hopping (between $B_1$-$A_2$), $-\gamma_3$ with sliding.
(e) The smallest inter-layer hopping (between $A_1$-$A_2$ or $B_1$-$B_2$), $-\gamma_4$ with sliding.
}
\end{figure}

Atomic structures of sliding BLG are shown in Figs. 1(a) and 1(b).
Without sliding, carbon atoms in BLG are arranged according to Bernal stacking - 
carbon atoms in one sublattice of the uppler layer are right 
on top of ones in the other sublattice of the lower layer.
When the bottom layer slides with respect to the top layer,
the sliding vector $\vec d_s$ can be written as a linear combination
of $\vec\delta_i$ ($i=1,2,3$)
which connect the nearest neighbor (nn) carbon atoms in the top layer
(For definitions of 
 $\vec\delta_i$ ($i=1,2,3$), see Fig. 1).
Considering the lattice structures of sliding bilayer,
it is easy to check that the sliding along $\pm\vec\delta_1$
is equivalent to one along $\pm\vec\delta_{2(3)}$.
We also note that a sliding along $\vec\delta_3$
is equivalent to one along $-\vec\delta_1$.
Full sliding along $\vec\delta_1$ brings an AB-stacking bilayer graphene
to a BA-stacking bilayer graphene while one along $\vec\delta_3$
shifts an AB-stacking bilayer to a AA-stacking bilayer.

\begin{figure}[t]
\includegraphics[width=1.0\columnwidth]{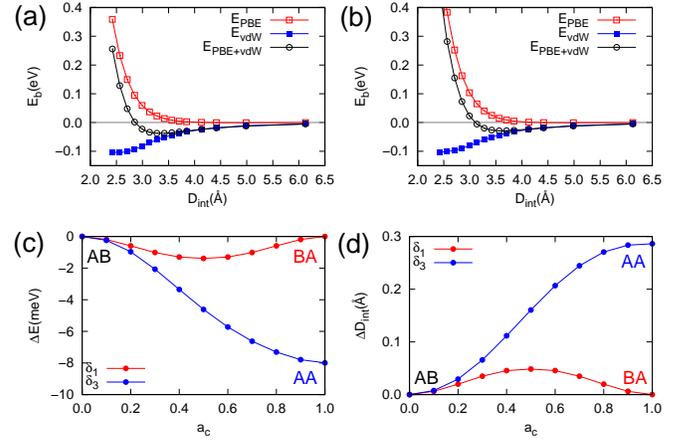}
\caption{(color online)
Interaction energy per atom ($E_b$) between two graphene layers
(a) with Bernal (AB) stacking and (b) with AA stacking
as a function of interlayer distance ($D_{\rm int}$).
$E_{\rm PBE}$, $E_{\rm vdW}$ and $E_{\rm PBE+vdW}$ denote
the interaction energies calculated
by PBE functional only, vdW correction only and including both.
Calculations results of $E_{\rm PBE}$ and $E_{\rm PBE+vdW}$ (retangles) are
fitted to a second-order Birch equation (solid lines).
$E_b=E_{\rm tot}-2e_{\rm tot}$ where $E_{\rm tot}$ is the total energy
of bilayer graphene per atom and $e_{\rm tot}$ that of single layer graphene per atom.
In (c) and (d), the variations of binding energy ($\Delta E$) and 
interlayer distance ($\Delta D_{\rm int}$)
with respect to those of AB-stacked bilayer are shown as functions of sliding distance in units of $a_c$
along $\vec\delta_1$ (red) and $\vec\delta_3$ (blue).
The negative $\Delta E$ denotes the decrease in binding energy.
We set the binding energy and interlayer distance of AB-stacked bilayer
to be zero in (a) and (b).
}
\end{figure}

We find that the equilibrium interlayer distance of bilayer graphene (BLG)
in Bernal stacking is 3.348\AA~and its binding energy
is $38.76$ meV (Fig. 2 (a)),
which are in good agreement with other calculations
and avaliable experimental data~\cite{barone,yang,dobson,galli,antony,langreth,crespi}.
When bilayer graphene has AA-stacking (one layer is right on top
of the other layer), the interlayer distance increases to 3.635\AA~
and binding energy decreases to $30.77$ meV (Fig. 2 (b)).
When one graphene layer slides with respect to the other along
either $\vec\delta_1$ or $\vec\delta_3$ direction,
the binding energy starts decreasing and interlayer distance increases
agreeing with a previous study~\cite{crespi} (Fig. 2 (c) and (d)).
We note that the change in the binding energy ($<$ 0.19 meV) and interlayer 
distance ($<$0.01 \AA) is quite negligible when the sliding distance is less than 0.14 \AA~
(about 10 \% of the bond length, $a_c$).

\section{Low energy band structures from GGA calculations}

\begin{figure}[t]
\includegraphics[width=1.0\columnwidth]{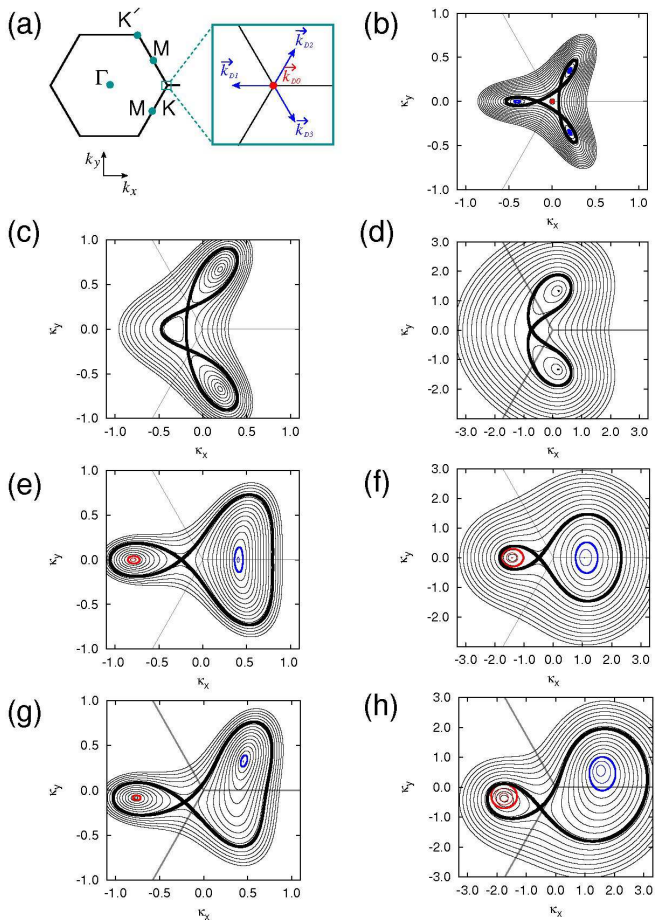}
\caption{
(a) The small rectangle near around K-point in the first Brillouin zone is enlarged 
to show the three-fold symmetric vectors $\vec k_{D1}=p_D(-1,0)$,
$\vec k_{D2}=p_D(1/2,\sqrt{3}/2)$, and $\vec k_{D3}=p_D(1/2,-\sqrt{3}/2)$
with respect to $\vec k_{D0}=0$, 
where $p_D\equiv (2/3a_c)(\gamma_1 \gamma_3/\gamma_0^2)$.  
(b) Energy contour for the valence bands of BLG without sliding. 
For all panels,   
$\kappa_x =100\times (k_x -\overline{\Gamma K})/\overline{\Gamma K}$
and $\kappa_y=100\times k_y /\overline{\Gamma K}$.
The thick contour is an iso-energy ($E = -1.1$ meV) curve crossing the three saddle points. 
The blue (red) contours denote the hole (electron) pockets. 
Energy contours for the valence band of BLG with sliding 
$\vec d_s =$ (c) $0.02 \vec\delta_1$,
(d) $0.1 \vec\delta_1$,
(e) $-0.02 \vec\delta_1$,
(f) $-0.1 \vec\delta_1$,
(g) $0.012 (\vec\delta_3-\vec\delta_1)$,
and 
(h) $0.1 (\vec\delta_3-\vec\delta_1)$. 
The contour interval is set by 0.5 meV for (a), by 1.0 meV for (c), (e) and (g) and by 10 
meV for (d), (f) and (h). 
The thick contours that cross the saddle points are at $E =$ (c) $-7.2$ meV, 
(d) $-32.7$ meV, (e) $-10.8$ meV, (f) $-48.5$ meV, (g) $-9.7$ meV 
and (h) $-74.0$ meV.
}
\end{figure}

Considering the low energy electronic structure of BLG in Bernal stacking, our 
calculations show, as in a previous study, four Dirac cones formed around 
the Fermi energy ($E_F$) at four Dirac points (${\vec k}_{Di}$ , $i = 0, 1, 2, 3$) 
near around $K$-point [Fig. 3(a) and 3(b)]~\cite{mccann,ando}. 
The magnitude of  ${\vec k}_{Di}$ ($i = 1, 2, 3$) is 
about 0.4\% of the distance between $\Gamma$- to $K$-point ($\overline{\Gamma K}$) [Fig. 3 (b)]. 
As the energy moves away from the $E_F$, 
the four Dirac cones merge to form three saddle points between the cones [Fig. 3(b)].
The calculation result can also be 
described by an effective Hamiltonian~\cite{mccann},   
\begin{equation}
{\mathcal H}_{\rm eff}(\vec k)=\hbar v_3 \vec\tau\cdot\vec k
+\frac{\hbar^2 v_0^2}{\gamma_1}(\vec\tau^*\cdot\vec k)\tau_x (\vec\tau^*\cdot\vec k),
\end{equation}
where $v_\alpha=(3/2)\gamma_\alpha a_c /\hbar$ ($\alpha=0,3)$, 
$\hbar \vec k=\hbar (k_x, k_y)$ is the crystal momentum from $K$-point,  
and $\vec\tau=(\tau_x,\tau_y)$ and $\vec \tau^*$ are Pauli spin matrices and their 
complex conjugates 
(See Fig. 1 for definition of $\gamma_\alpha$).
Typical estimates are $\gamma_0\simeq 3$ eV, $\gamma_1\simeq\gamma_3\simeq\gamma_0/10$
and $\gamma_4 \simeq \gamma_0/20$ (Ref.~\cite{kuzmenko}). 
We will neglect $\gamma_4$ and discuss its role in Sec. VI. 
If we expand the effective Hamiltonian around each $\vec k_{Di}$,
we have one isotropic Dirac Hamiltonian,
${\mathcal H}_{D0}=\hbar v_3 \vec\tau\cdot\vec{\delta k}_0$
at $\vec k_{D0}$, an anisotropic
${\mathcal H}_{D1}=-\hbar v_3 (\tau^*_x \delta k_{x1}+3\tau^*_y \delta k_{y1})$
at $\vec k_{D1}$, and two anisotropic others at $\vec k_{D2}$ and $\vec k_{D3}$
which can be obtained by rotating $H_{D1}$ by $\pm2\pi/3$ respectively.
Here, $\vec{\delta k}_i=(\delta k_{xi}, \delta k_{yi})=\vec k-\vec k_{Di}$, $(i=0,1,2,3)$.

We find that the low energy bands of BLG change dramatically when one layer slides 
with respect to the other in an extremely small amount and in arbitrary directions. 
Let the bottom layer slide with respect to the top by ${\vec d}_s(=(d_x,d_y))$  [Fig. 1(a) and 1(b)]. 
First, when the bottom layer moves along ${\vec\delta}_1$ 
direction by 0.028 \AA, i.e., ${\vec d}_s=0.02 {\vec\delta}_1$, 
only two Dirac cones at ${\vec k}_{D2}$ and ${\vec k}_{D3}$ remain, 
instead of four cones for BLG without sliding as shown in  Fig. 3(c)
The energetic position of the saddle points of the valence band 
decreases from $-1.1$ meV to $-7.2$ meV. 
When the bottom layer slides further (${\vec d}_s=0.1 {\vec\delta}_1$), 
the topology changes again and the saddle point energy significantly decreases to $-32.7$ meV [Fig. 3(d)]. 
Second, when the bottom layer 
moves along the direction opposite to the previous one (${\vec d}_s=-0.02 {\vec\delta}_1$), 
the low energy bands changes again completely. 
In this case, the Dirac cone at ${\vec k}_{D1}$ moves along the $-k_x$ 
direction and an anomalous Dirac cone with sickle-shaped energy contours appears at a 
new Dirac point instead of three cones [Fig. 3(e)]. 
If the bottom layer slides further by 
0.14 \AA~along $-{\vec\delta}_1$
(${\vec d}_s=-0.1 {\vec\delta}_1$), 
the topology remains the same and the anomalous cone 
comes to have an anisotropic shape [Fig. 3(f)]. In this case, the saddle point energy 
decreases to $-48.5$ meV (almost 500\% of pristine one) [Fig. 3(f)]. 
Finally, when we slide the bottom layer by $0.012({\vec\delta}_3-{\vec\delta}_1)$ [Fig. 3(g)] 
and $0.1({\vec\delta}_3-{\vec\delta}_1)$ [Fig. 3(h)], the topological 
changes are similar to the case for the sliding along $-{\vec\delta}_1$. 
The saddle point energy for the 
sliding by $0.1({\vec\delta}_3-{\vec\delta}_1)$ decreases dramatically down to $-$74.0 meV as shown in Fig. 3(h). 
After a comprehensive search for the topological changes by the sliding along arbitrary 
directions (not shown here), we find that the topology of saddle point energy 
contours is all similar to those shown in Fig. 3(e)-(h) (one crossing point) except the 
topologically distinctive phase in Fig. 3(c) (two crossing points).

\section{Effective Model Hamiltonians and non-abelian gauge potential}

The hypersensitive topological changes found by first-principles calculations 
in the previous section demonstrate 
that the sliding motion creates interactions between the effective non-Abelian 
SU(4) gauge field background and massless fermions. 
In the presence of a very tiny sliding, the nnn interlayer interaction ($\gamma_3$) 
is not constant any more but depends exponentially on the different pair distances
between carbon atoms in top and bottom layers as shown in Fig. 1(d). 
The asymmetric inter-layer hopping interaction produces a 
constant pseudo-gauge potential in the terms of Hamiltonian containing $\gamma_3$ only,
being similar with an effective Hamiltonian for strained SLG~\cite{pereira,choi,guinea2,choi2}. 
Hence, the effective Hamiltonian for sliding BLG can be written as, 
\begin{equation}
{\mathcal H}'_{\rm eff}(\vec k)=\hbar v_3 \vec\tau\cdot(\vec k -\vec \lambda)
+\frac{\hbar v_3}{p_D}(\vec\tau^*\cdot\vec k)\tau_x (\vec\tau^*\cdot\vec k)
\end{equation}
where $p_D\equiv\gamma'_1 v_3 /(\hbar v_0^2)$ and $\gamma'_1$ is the reduced nn inter-layer interaction.
(See Appendix for derivation of Eq. (2)).
We find that the constant vector potential ($\vec\lambda$) explicitly depends on the sliding vector, 
$\vec\lambda\equiv (\lambda_x,\lambda_y)=\beta(d_y,-d_x)$.
By fitting the energy bands obtained by the model to ones by the first-principles calculation,
we can estimate that $\beta$ is about $1/a_c^2$.
The gauge symmetry in Eq. (2) apparently seems to be broken since 
$\vec\lambda$ is absent in the quadratic term of $\vec k$. 
However, when we expand Eq. (1) at each Dirac point ($\vec k_{Di}$), 
the four Dirac cones shift depending on both 
$\vec\lambda$ and their own positions as $\vec k_{Di}\rightarrow \vec k_{Di}+\vec A_i(\vec\lambda)$ ($i = 0, 1, 2, 3$). 
If $|\vec \lambda|\ll p_D$, $\vec A_0 = (\lambda_x, \lambda_y)$, $\vec A_1 = (-\lambda_x,  \frac{\lambda_y}{3})$,
$\vec A_2 =(-\frac{\sqrt{3}\lambda_y}{3} , -\frac{\sqrt{3}\lambda_x}{3}  - \frac{2 \lambda_y}{3} )$
and $\vec A_3 =(\frac{\sqrt{3}\lambda_y}{3} , \frac{\sqrt{3}\lambda_x}{3}  - \frac{2\lambda_y}{3} )$.
So, each Dirac cone moves along a different direction depending on its position. 

\begin{figure*}[t]
\includegraphics[width=2.0\columnwidth]{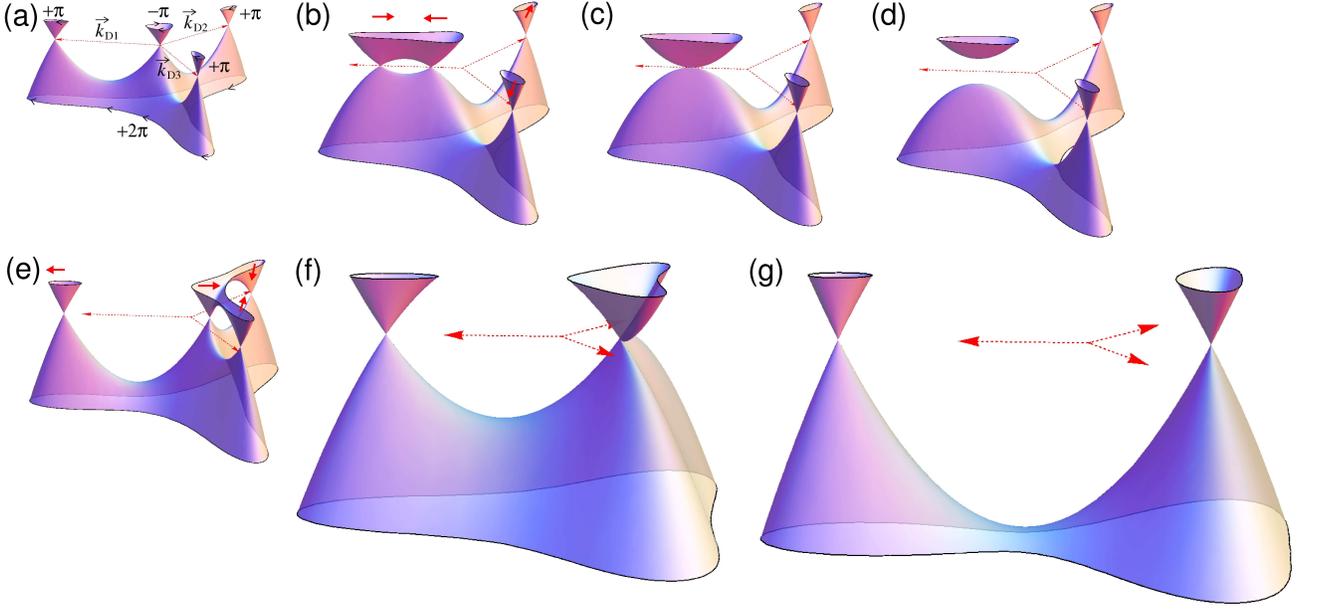}
\caption{
Three dimensional plots of low 
energy bands from the effective Hamiltonian of Eq. (1)
with $\vec\lambda=(\lambda_x,0)$. 
(a) Without sliding ($\vec \lambda=0$), 
three anisotropic Dirac cones with Berry phase $\pi$ 
at $\vec k_{Di}$ ($i=1, 2, 3$) denoted by dotted red arrows in all panels,
and isotropic one with $-\pi$ at $\vec k_{D0}$. 
Energy bands for sliding BLG with (b) $\lambda_x=-p_D/5$ 
(c) $-p_D/4$, (d) $-p_D/3$, (e) $p_D/5$, (f) $3p_D/4$
and (g) $1.8 p_D$.
In term of sliding distance $d_y$,
(b) $d_y\simeq -0.003~$,~(c) $-0.004$,~(d) $-0.005$,~(e) $+0.003$,~
(f) $+0.012$ and (g) $+0.029$ \AA.
The solid red arrows in (b) and (e) indicate the movements of Dirac cones at sliding.
}
\end{figure*}

The essential feature of the system that enables the electronic topological transition in BLG
is the shift of four Dirac points along different directions under the sliding 
of one layer against the other.
As we have shown, the effect of sliding is to replace $\vec{\delta k}_i=\vec{k}-\vec{k}_{Di}$ 
with $\vec{\delta k}_i-\vec{A}_i$ in the effective Hamiltonian 
for the low energy modes near the Dirac points and this feature suggests that 
the sliding in BLG induces a non-Abelian background gauge field associted with these modes.
In order to show this more explicitly, we assume that before sliding, 
there are four massless Dirac fermions, labeled by $i=0,1,2,3$, whose Hamiltonians are all isotropic and of same chirality:
\begin{equation}\label{symmham}
\mathcal{H}_{Di}=\hbar v_3 \vec{\tau}^*\cdot\vec{\delta k}_i\,,\quad i=0,\ldots,3.
\end{equation}
We can then form a quadruplet out of four fermions, 
so that the index $i$ is now viewed as the ``color'' index of SU(4) symmetry, 
and combine the four Hamiltonians $\mathcal{H}_{Di}$ compactly as 
$\mathcal{H}_{D}=\hbar v_3 (\vec{\tau}^*\cdot\vec{\delta k})\otimes\mathbf{I}$, 
where $\mathbf{I}$ is the $4\times 4$ identity matrix. 
If we now introduce the background gauge field $\vec{\mathbf{A}}=(\mathbf{A}_x,\mathbf{A}_y)$ 
for the SU(4) symmetry with
\begin{equation*}
\mathbf{A}_x=\lambda_x
\begin{pmatrix}
1 &  &  &  \\
 & -1 &  &  \\
 &  & 0 &  \\
 &  &  & 0
\end{pmatrix}
-\frac{\lambda_y}{\sqrt{3}}
\begin{pmatrix}
0 &  &  &  \\
 & 0 &  &  \\
 &  & 1 &  \\
 &  &  & -1
\end{pmatrix}\,,
\end{equation*}
and
\begin{equation*}
\mathbf{A}_y=\frac{\lambda_x}{\sqrt{3}}
\begin{pmatrix}
0 &  &  &  \\
 & 0 &  &  \\
 &  & -1 &  \\
 &  &  & 1
\end{pmatrix}
+\lambda_y
\begin{pmatrix}
1 &  &  &  \\
 & \tfrac13 &  &  \\
 &  & -\tfrac23 &  \\
 &  &  & -\tfrac23
\end{pmatrix}\,,
\end{equation*}
so that $\vec{\delta k}\mathbf{I}\rightarrow\vec{\delta k}\mathbf{I}-\vec{\mathbf{A}}$ 
in the Hamiltonian, then Eq. (\ref{symmham}) now becomes
\begin{equation}\label{symmham2}
\mathcal{H}_{Di}=\hbar v_3 \vec{\tau}^*\cdot(\vec{\delta k}_i-\vec{A}_i)\,,\quad i=0,\ldots,3,
\end{equation}
where $\vec{A}_i$ are precisely the shifts of the Dirac points shown 
at the beginning of this section. 
Since both $\mathbf{A}_x$ and $\mathbf{A}_y$ are linear combinations 
of the generators of SU(4), we now see the effect of sliding 
as if introducing a non-Abelian background gauge field.

The low energy Hamiltonian of bilayer graphene is different from what we have just shown above 
in that $\mathcal{H}_{D0}$ has opposite chirality from $\mathcal{H}_{Di}$ for $i=1,2,3$, and that 
the latter three Hamiltonians are anisotropic. 
A parity inversion $\delta k_y\rightarrow - \delta k_y$ for $\mathcal{H}_{D0}$ 
and anisotropic rescalings of $\vec{\delta k}_i$ for $\mathcal{H}_{Di}$ $(i=1,2,3)$
transform SU(4) isotropic quadruplet to the low energy Hamiltonian of BLG.
This transformation is equivalent to multiplying
the Pauli matrices $\tau_x$ and $\tau_y$ with (different) constants 
in Eq. (\ref{symmham2}) --- for example, $(\tau_x,  \tau_y)\rightarrow (\tau_x,-\tau_y)$ 
for $i=0$, and $(\tau_x,  \tau_y)\rightarrow -(\tau_x,3\tau_y)$ for $i=1$ --- thereby 
breaking the SU(4) symmetry of the previous paragraph. 
But the structure of the Hamiltonians otherwise remains the same, and in particular,
the shifts of the Dirac points are still given by $\vec{A}_i$ of Eq. (\ref{symmham2}).  

\section{Roles of Berry's phase in electronic topological transition}

The characteristics of the ETT are ruled by unique interplay between the
effective non-Abelian vector potential and conservation of Berry's phase. 
The Berry's phase ($\phi_B$) for each Dirac cone at $\vec k_{Di}$ can be calculated by
using $\phi_B=\oint_\Gamma d\vec k \cdot {\mathcal A}(\vec k)$
where ${\mathcal A}(\vec k)=i\langle u_{Di}(\vec k)|\partial_{\vec R}|u_{Di}(\vec k)\rangle$,
$\Gamma$ is a path enclosing each Dirac point, and $u_{Di}(\vec k)$ is
a single-valued spinor-like eigenfunction of each Dirac Hamiltonian at $\vec k_{Di}$.
Without sliding, we have $\phi_B=+\pi$ for the massless Dirac fermions around $\vec k_{Di}$ ($i=1,2,3$)
	and $\phi_B=-\pi$  for the ones at the center $\vec k_{D0}$ [Fig. 4(a)].
For a path enclosing all Dirac points at higher energy,
we can use the quadratic Hamiltonian (1)
and $\phi_B$ is $+2\pi$~\cite{novoselov,lemonik}.
So, the total $\phi_B$ is always conserved to be $2\pi$ [Fig. 4(a)]~\cite{lemonik}.

Now, to reveal the role of Berry's phase explicitly, 
let us consider exactly solvable cases without
the assumption of $|\vec \lambda|\ll p_D$.
For sliding along $\pm\vec\delta_1$ direction, i.e., $\vec d_s =(0,\mp d_y)$ ($d_y>0$),
the effective vector potential is given by
$\vec \lambda =(\lambda_x,0)=(\mp\beta d_y,0)$.
Here, $\lambda_x<0~(\lambda_x >0)$ when sliding along $+\vec\delta_1 (-\vec\delta_1)$ direction.
Then we have four different local vector potentials $\vec A_i$ for each $\vec k_{Di}$
such that
\begin{eqnarray*}
\vec A_0&=&\frac{p_D}{2}\left(-1+\sqrt{1+\frac{4\lambda_x}{p_D}},0\right),\nonumber\\
\vec A_1&=&\frac{p_D}{2}\left(+1-\sqrt{1+\frac{4\lambda_x}{p_D}},0\right),\nonumber\\
\vec A_2&=&\frac{\sqrt{3}p_D}{2}\left(0,-1+\sqrt{1- \frac{4\lambda_x}{3p_D}}\right),\nonumber\\
\vec A_3&=&\frac{\sqrt{3}p_D}{2}\left(0,+1-\sqrt{1- \frac{4\lambda_x}{3p_D}}\right).
\end{eqnarray*}
We note that $\vec A_0+\vec A_1=0$ and $\vec A_2+\vec A_3=0$.
Thus, the Dirac cone at $\vec k_{D0}$ and the one at $\vec k_{D1}$
move in opposite direction when bilayer graphene slides along $\pm\vec\delta_1$
and so do those at $\vec k_{D2}$ and $\vec k_{D3}$.
It is also noticeable that $\vec k_{D0}+\vec A_0 = \vec k_{D1}+\vec A_1$ when $\lambda_x=-p_D/4$
and $\vec k_{D1}+\vec A_1 = \vec k_{D2}+\vec A_2= \vec k_{D3}+\vec A_3$  when $\lambda_x=3p_D/4$.
Hence, when the bottom layer slides along $-y$ direction ($\vec\delta_1$ direction)
by $d_y=p_D/(4\beta)$,
the two Dirac cones at $\vec k_{D0}$ and $\vec k_{D1}$
meet at $(-p_D/2,0)$
while three cones at $\vec k_{D1}$, $\vec k_{D2}$ and $\vec k_{D3}$
meet together at $(p_D/2,0)$ when sliding along $+y$ direction ($-\vec\delta_1$ direction)
by $d_y=3p_D/(4\beta)$.

For the sliding along $+\vec\delta_1$ direction, the effective Hamiltonian shown in Eq. (2)
can be expanded around $\vec k_+=(-p_D/2,0)$ so that the resulting Hamiltonian can be written as
\begin{equation}
{\mathcal H}_+\simeq \hbar v_3 \tau_x \left[\frac{(\delta k_x)^2}{p_D}-\left(\lambda_x+\frac{p_D}{4}\right)\right]
+2\hbar v_3 \tau_y \delta k_y
\end{equation}
where $\vec{\delta k}=(\delta k_x, \delta k_y)\equiv \vec k-\vec k_+$ and $\lambda_x <0$.
The new effective Hamiltonian (5) has eigenvalues given by
\begin{equation}
E_+(\vec{\delta k})=\pm\hbar v_3\sqrt{\left[\frac{(\delta k_x)^2}{p_D}
	-\left(\lambda_x+\frac{p_D}{4}\right)\right]^2+4(\delta k_y)^2}.
\end{equation}
So, when $-p_D/4<\lambda_x <0$ or $0<d_y<p_D/(4\beta)$, 
there are still two Dirac cones at $\vec k_{D0}+\vec A_0$ and $\vec k_{D1}+\vec A_1$.
When $\lambda_x$ reaches to a critical value of $-p_D/4$,
the Hamiltonian is given by
$\hbar v_3 \tau_x (\delta k_x)^2/p_D +2\hbar v_3 \tau_y \delta k_y$
so that the dispersion along $k_x$ direction
is massive while one along $k_y$ direction is still massless.
Two other cones at $\vec k_{D2}+\vec A_2$
and $\vec k_{D3}+\vec A_3$ move away from each other in $\pm k_y$ direction
maintaining their anisotropic Dirac cone shapes.
Therefore, it can be seen that the cones at $\vec k_{D0}$ and $\vec k_{D1}$  merge together  
when $\vec\lambda=(-p_D/4,0)$ [Fig. 4(c)].
The corresponding sliding distance of $d_y$  
is about 0.3\% of $a_c$ ($d_y=p_D/(4\beta)\sim 0.004$\AA).  
The low energy bands already change their topology 
from the original structure under extremely small sliding. 
For further sliding, $\lambda_x < -p_D/4$ ($d_y>p_D/(4\beta)$), the merged 
cone eventually disappears and the spectrum of Eq. (6) develops
an energy gap at ${\vec k}_+$ as shown in Fig. 4(d). 
The gap is linear with sliding 
distance as given by $2\hbar v_3|\lambda_x+p_D/4|\simeq 2\hbar v_3 \beta d_y \sim 0.9\times [d_y /a_c]$ eV. 
The opening of energy gap signals {\it a pair annihilation of two massless 
Dirac electrons} with the opposite `topological charges' of $\pm 1$
since the two Dirac cones at $\vec k_{D0}$ and $\vec k_{D1}$ have the Berry's phase 
of $\pm\pi$ respectively. 
This also confirms the Berry's phase conservation since 
the remaining two anisotropic Dirac cones at $\vec k_{D2}$ and $\vec k_{D3}$ give the total $\phi_B$ of $2\pi$.

Next, for the sliding along $-\vec\delta_1$ direction, the effective Hamiltonian
can be obtained by expanding Eq. (2) around $\vec k_-=(p_D/2,0)$:
\begin{eqnarray}
{\mathcal H}_-&\simeq& 
\hbar v_3 \tau_x^*
\left[2\delta k_x +\frac{(\delta k_x)^2-(\delta k_y)^2}{p_D}
+\left(\frac{3p_D}{4}-\lambda_x\right)\right] \nonumber\\
& &+2\hbar v_3 \tau_y^* \frac{(\delta k_x)(\delta k_y)}{p_D},
\end{eqnarray}
where $\vec{\delta k}=(\delta k_x, \delta k_y)\equiv \vec k-\vec k_-$ and $\lambda_x >0$.
When $0<\lambda_x <3p_D/4$, there are three Dirac cones at $\vec k_{Di}+\vec A_i$ ($i=0,2,3$).
However, when $\lambda_x=3p_D/4$, the three Dirac cones merge
at $\vec k_-$ and this Hamiltonian gives an anomalous
dispersion relation written by
$E_-(\vec{\delta k})=
\pm\hbar v_3\sqrt{f^2(\vec{\delta k})+g^2(\vec{\delta k})}
$ where
$f(\vec{\delta k})=2\delta k_x +(\delta k_x)^2/p_D-(\delta k_y)^2/{p_D}$
and $g(\vec{\delta k})=2{(\delta k_x)(\delta k_y)}/{p_D}$.
This gives sickle-shaped constant energy contours which are consistent with 
our {\it ab initio} calculation results shown in Figs. 3(e) and (g).
When $\lambda_x > 3p_D/4$, the effective Hamiltonian (7) 
does not develop any energy gap at all.
Instead, when $\lambda_x \gg 3p_D/4$,
the effective Hamiltonian has a Dirac point at $\vec k'_-=(-p_D/2+\sqrt{\lambda_x p_D},0)$
and is given by
\begin{equation}
{\mathcal H}'_-\simeq 2\hbar v_3 \sqrt{\frac{\lambda_x}{p_D}}\tau^*_x\delta k_x
+2\hbar v_3 \left(\sqrt{\frac{\lambda_x}{p_D}} -1\right)\tau^*_y\delta k_y,
\end{equation}
where
$\vec{\delta k}=(\delta k_x, \delta k_y)\equiv \vec k-\vec k'_-$.
So, as sliding distance increases along $-\vec\delta_1$,
the anomalous Hamiltonian (7) gradually transforms to
the anisotropic Dirac Hamiltonian (8).
This is quite contrary to the gapped spectrum (6) generated
by sliding motion along $\vec\delta_1$ direction ($\mathcal H_+$ in
Eq. (5)).
The other cone at $\vec k_{D1}+\vec A_1$ moves in $-k_x$ direction
maintaining its anisotropic dispersion relation.
As $\lambda_x\gg 3p_D/4$, the Dirac cone at $\vec k_{D1}+\vec A_1$
has an asymtotic shape as following,
\begin{equation}
{\mathcal H}''_-
\simeq -2\hbar v_3 \sqrt{\frac{\lambda_x}{p_D}}\tau^*_x\delta k_x
-2\hbar v_3 \left(\sqrt{\frac{\lambda_x}{p_D}} +1\right)\tau^*_y\delta k_y,
\end{equation}
where
$\vec{\delta k}=(\delta k_x, \delta k_y)\equiv \vec k-\vec k_{D1}-\vec A_1$.
Therefore, in contrast to the first case, no energy gap develops even when the sliding distance is 
increased further. Instead, the anomalous dispersion transforms to an anisotropic Dirac 
cone [Fig. 4(g)]. This phenomenon can be interpreted as a merging of two massless 
fermions of topological charge $+1$ with one of topological charge $-1$. 
As a result, {\it a new fermion of topological charge $+1$ is generated}. 
We note that the total $\phi_B$ of $2\pi$ is 
conserved since the new particle has $\phi_B$ of $\pi$. Hence, the topological charges of 
fermionic particles in BLG are strictly governed by Berry's phase conservation rule.

\section{Effect of smallest interlayer hoppings}

When sliding occurs, the smallest interlayer interaction $\gamma_4$ shown in Fig. 1(e) 
becomes anisotropic and depends on the pair distances 
between relevant carbon atoms in the top and bottom layer.
This effect adds an additional Hamiltonian to Eq. (2),
\begin{eqnarray}
{\mathcal H}'_{\rm eff}&\simeq& \frac{2\hbar^2 v_0 v_4}{\gamma_1}\vec k\cdot(\vec k+\vec\lambda) \tau_0 \nonumber \\
& + &\frac{\hbar^2 v^2_4}{\gamma_1}[\vec\tau^* \cdot (\vec k+\vec \lambda)]\tau_x [\vec\tau^*\cdot(\vec k+\vec\lambda)],
\end{eqnarray}
where $v_4=\frac{3\gamma_4 a_c}{2\hbar}$
(See Appendix for derivation of Eq. (10)).
Since the second term in Eq. (10) is twenty times smaller than the first term, we will neglect the second term hereafter.
When sliding along $\pm\vec\delta_1$, the difference between Dirac energies at $\vec k_{D1}$ and $\vec k_{D0}$ is
given by $\frac{2\gamma_1 v_3^2 v_4}{v_0^3}(1-\frac{\lambda_x}{p_D})\sqrt{1+\frac{4\lambda_x}{p_D}}$.
Without sliding ($\lambda_x=0$), the difference becomes $\Delta\varepsilon_{\rm eh}=\frac{2\gamma_1 v_3^2 v_4}{v_0^3}$,
which indicates the hole and electron doped Dirac cone at $\vec k_{D1}$ and $\vec k_{D0}$ respectively.
This explains the hole and electron doped cones shown in our {\it ab initio} calculation results [Fig. 3(b)].
With sliding along $\vec\delta_1$, the difference disappears when
 $\lambda_x<-p_D/4$ as shown in Figs. 3(c) and 3(d) so that all Dirac cones are charge-neutral.
 Contrary to this, when sliding along $-\vec\delta_1$, the difference changes its sign
 when $\lambda_x>p_D$ and decreases significantly as sliding distance increases.
So, the hole-doped Dirac cone at $\vec k_{D1}$
 changes to be electron-doped and the new Dirac cone to be hole-doped with increasing amount of
 doping as increasing sliding distance as shown in our first-principles calculations [Figs. 3 (e)-(h)].
We note that variations in $\gamma_4$ do not affect any topological changes discussed so far. 

\begin{figure}[t]
\includegraphics[width=1.0\columnwidth]{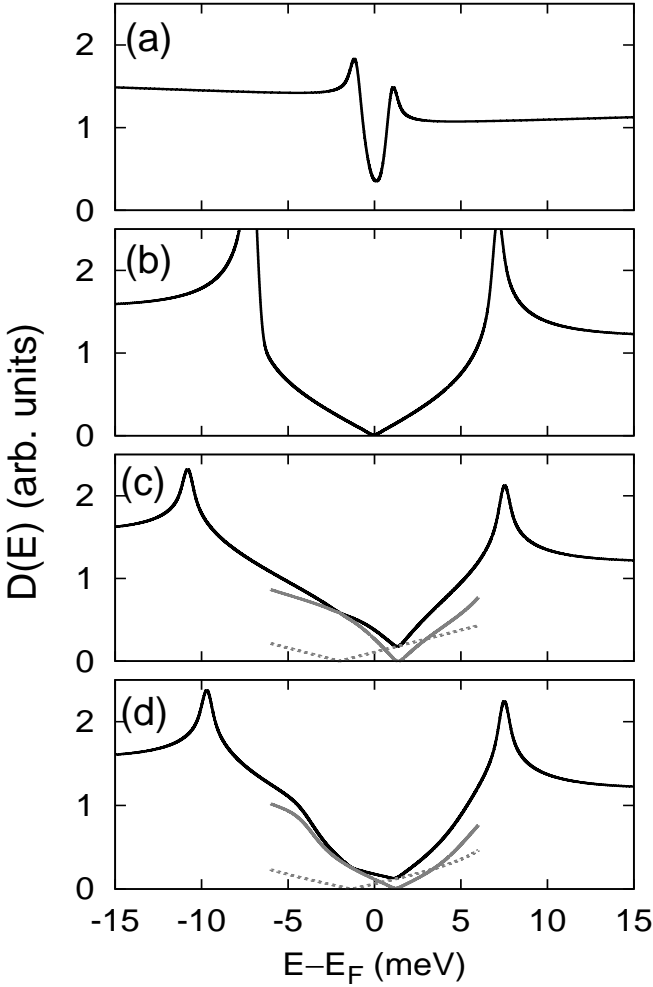}
\caption{
First-principles calculations of total density of states 
of sliding BLG when $\vec d_s=$ (a) 0, (b) $0.02\vec\delta_1$, 
(c) $-0.02\vec\delta_1$, and (d) $0.012(\vec\delta_3-\vec\delta_1)$, respectively.
The grey solid and dotted lines in (c) and (d) are 
the partial density of states projected onto $\kappa_x > \kappa_s$ and $\kappa_x < \kappa_s$ region 
of the BZ in Fig. 3(e) and 3(g), respectively, where $\kappa_s$ is a $x$-component 
of the saddle point of each energy contour.
}
\end{figure}

\section{Discussion}

The direct signatures of the ETT can be readily measured using 
various experiment methods. 
We showed that the tiny sliding lifts the degeneracy at $E_F$ 
as the number of Dirac cones is always reduced from four to two
with significant increase of the saddle point energies and deformations of remaining Dirac cones.
Therefore, first, high-resolution scanning tunneling microscopy~\cite{li,song} can directly measure the 
changes in the density of states ($D(E)$). 
Between saddle point energies $|E|<\varepsilon_s$ where $\varepsilon_s=\gamma_1 v_3^2/(4v_0^2)$
without sliding,
the total density of states per unit area is given by $D(E)=|E|/(\pi v_3^2)$.
When BLG slides along $\pm\vec\delta_1$, saddle point energies increase linearly as
sliding distance increases, $\varepsilon_s (d_y)=\hbar v_3 |\lambda_x -p_D/4|\sim \hbar v_3 \beta d_y$.
When sliding along $\vec\delta_1$, the two Dirac cones disappear quickly and the other two remaining ones
are anisotropic massless Dirac cones. So, the density of states in between saddle point energies $\pm\varepsilon_s(d_y)$
linearly depends on energy, $D(E)\sim |E|$. 
Our first-principles calculations 
indeed show drastic variations in the position of van Hove singularity (vHS) below and 
above the charge neutral point upon sliding (Fig. 5). 
In between the two vHSs, a linear [Figs. 5(a) and 5(b)] or a mostly square-root dependence 
of $D(E)$ [Figs. 5(c) and (d)] appears as the sliding direction is changed, which is a unique 
feature of two-dimensional materials~\cite{goerbig,wunsch}. 

Second, the Landau level (LL) spectrum for a 
small perpendicular magnetic field ($B$) also exhibits distinctive dependence on the sliding 
direction. 
By using semiclassical quantization rule under the perpendicular magnetic fields ($B$), 
$S(E)=\frac{2\pi |e|B}{\hbar c}(n+\gamma)$
and $\partial_E S(E)=4\pi^2 D(E)$ where $S(E)$ is an area of closed orbit of electron 
and $\gamma=1/2-\phi_B/2\pi$~\cite{mikitik,goerbig,dietl}, 
we can immediately confirm that the Landau level (LL) spectrum under a small magnetic field is given
by $E_n \sim \pm(Bn)^{1/2}$ in the case of sliding BLG along $\vec d_s=0.02\vec\delta_1$. 
When BLG slides by either $-0.02\vec\delta_1$ or $0.012(\vec\delta_3-\vec\delta_1)$, 
the anomalous Dirac cone shape results in $D(E)\sim \sqrt{E}$ being similar to the previous study on
the density of states of semi-Dirac cone (massive in one direction and massless in the other)~\cite{goerbig,dietl}.
By using the rule above and $D(E)\sim \sqrt{E}$, the LL spectrum is given by $E_n\sim\pm(Bn)^{2/3}$.
It is noticeable that, irrespective of sliding direction, the zeroth LL exists at zero energy since the topological
charge conservation (Berry's phase conservation) enforce the existence of at least one massless modes in the system.

Third, an ideal minimal conductivity of $24e^2/(\pi^2\hbar)$ at the charge neutral 
point~\cite{ando,nilsson,cserti} will decrease quickly when sliding occurs ($e$ is the electron charge). 
The conductivity of anisotropic massless Dirac fermions with a dispersion, $v_x \tau_x k_x+v_y \tau_y k_y$
is given by $\sigma=g\frac{e^2}{\pi^2 \hbar}\frac{v_x}{v_y}$~\cite{nilsson} where $g$ is a degeneracy
factor ($g=4$ if considering valley and spin degeneracies) and $e$ is an electron charge.
So, by using Eqs (8) and (9), the conductivity in wide BLG sliding along $-\vec\delta_1 ~(d_y\gg p_D)$
can be calculated easily, 
$\sigma=\frac{8e^2}{\pi^2 \hbar}\left(1-p_D/\lambda_x\right)^{-1}
\simeq\frac{8e^2}{\pi^2 \hbar}\left(1+p_D/\lambda_x\right)$.
The conductivity approaches $8e^2/(\pi^2 \hbar)$ as sliding distance increases.

In summary, we show that the topology of energy bands of BLG changes significantly
if sliding of extremely small distance occurs in any direction. 
The effective non-Abelian background gauge potential can be generated by sliding 
motion and is shown to play an important role in dictating the characteristics
of sliding induced ETT. 
Hence, the ETT driven by non-Abelian gauge fields that are thought to be 
possible in cold atomic gas~\cite{goerbig,wunsch,bermudez} or 
similar effects in high energy physics~\cite{hosotani} can be realizable in sliding BLG.

{\it Note added in proof} After submission, we became aware of related works on similar systems from 
other groups~\cite{marcin,gail,mayorov}

\section*{Acknowledgements}
Y.-W.S. acknowledges discussions with K. Lee, P. Yi, 
and K. Novoselov. Y.-W.S. was supported by the NRF grant funded by the 
Korea government (MEST) (QMMRC, No. R11-2008-053-01002-0 and 
Nano R\&D program 2008-03670). S.-M.C. and S.-H.J. were supported by NRF funded by MEST (Grant 
2009-0087731 and WCU program No. R31-2008-000-10059-0).
We thank KIAS for providing computing resources (KIAS CAC Linux Cluster System).

\appendix*\section{Derivation of Eqs. (2) and (10)}

A single particle Hamiltonian of BLG in Bernal stacking,
  $H=H_0+H_1+H_2$, can be written as
\begin{eqnarray}
H_0
&=&-\gamma_0 \sum_{{\vec r},j,\alpha}
a_\alpha^\dagger({\vec r})
b_\alpha({\vec r}+{\vec \delta}_j)+{\rm (h.c.)},\\
H_1
&=&
-\gamma_1 \sum_{{\vec r}}a_1^\dagger ({\vec r})b_2 ({\vec r}) \nonumber\\
& &-\gamma_3\sum_{{\vec r},j} a_2^\dagger({\vec r}+{\vec \delta}_j)b_1({\vec r})
+{\rm (h.c.)},\\
H_2
&=&
-\gamma_4\sum_{{\vec r},j} a_1^\dagger({\vec r}+{\vec \delta}_j)a_2({\vec r})\nonumber\\
& &-\gamma_4\sum_{{\vec r},j} b_1^\dagger({\vec r}+{\vec \delta}_j)b_2({\vec r})
+{\rm (h.c.)},
\end{eqnarray}
where $H_0$ is a Hamiltonian for intra-layer interactions in each SLG and
$H_1$ and $H_2$ are for inter-layer interactions between two SLG.
Here $a^\dagger_\alpha (a_\alpha)$ and $b^\dagger_\alpha (b_\alpha)$ are
the creation (annihilaton) operator
of $\pi$-electron located at $A$- and $B$-sublattice of layer $\alpha~(=1,2)$ respectively,
$\vec r=m\vec a_1+n\vec a_2$ ($m$ and $n$ are integers),
$\vec a_1(=\vec\delta_3-\vec\delta_1)$
and $\vec a_2(=\vec\delta_2-\vec\delta_1)$ are unit vectors
of hexagonal lattice of SLG~\cite{mccann,ando,novoselov,neto,nilsson}.
$\gamma_0$ is the intra-layer nn hopping parameter,
and $\gamma_1$ and $\gamma_3$ are for the nn
and next nn (nnn) inter-layer hoppings parameters, 
respectively~\cite{mccann,ando,novoselov,neto,nilsson}.

Using Fourier transformations of
$a_\alpha(\vec r)
=\frac{1}{\sqrt{N}}\sum_{\vec p} e^{-i\vec p \cdot \vec r}a_{\alpha\vec p}$
and
$b_\alpha(\vec r)
=\frac{1}{\sqrt{N}}\sum_{\vec p} e^{-i\vec p \cdot \vec r}b_{\alpha\vec p}$
($N$ is a total number of unitcells),
the total Hamiltonian ($H$) in Eqs. (A.1)-(A.3) 
can be written in a matrix form such as
$H=\sum_{\vec p}\Psi_{\vec p}^\dagger {\mathcal H}_{\vec p}\Psi_{\vec p}$
for a field of $\Psi_{\vec p}=(b_{1\vec p}, a_{2\vec p}, a_{1\vec p}, b_{2\vec p})^T$
where
\begin{equation}
{\mathcal H}_{\vec p}=
\begin{pmatrix}
0       & \xi_3^*(\vec p) &\xi_0(\vec p) &\xi_4(\vec p) \\
\xi_3(\vec p)  &   0   & \xi_4^*(\vec p) &\xi_0^*(\vec p) \\
\xi_0^*(\vec p)&\xi_4(\vec p) & 0 & -\gamma_1 \\
\xi_4^*(\vec p)&\xi_0(\vec p) & -\gamma_1 & 0
\end{pmatrix},
\end{equation}
and $\xi_\alpha(\vec p)=-\gamma_\alpha\sum_j e^{i\vec p\cdot\vec\delta_j}$
$(\alpha=0,3,4,~j=1,2,3)$.
When we expand Eq. (A.4) around $K$-point 
by using $\vec p =\vec k+\vec K$ ($|\vec k|\ll |\vec K|$)
and $\vec K =(\frac{4\pi}{3\sqrt{3}}\frac{1}{a_c},0)$,
\begin{eqnarray}
\xi_\alpha(\vec k+\vec K)&\simeq&-\gamma_\alpha\sum_j e^{i\vec K\cdot\vec\delta_j}
-i\gamma_\alpha\sum_j \vec k\cdot\vec\delta_j e^{i\vec K\cdot\vec\delta_j}\nonumber\\
&=&v_\alpha(k_x+ik_y),	
\end{eqnarray}
where  $v_\alpha=\frac{3}{2\hbar}\gamma_\alpha a_c~(\alpha=0,3,4)$,
	   $\hbar$ is the Planck constant and
	  $\hbar{\vec k}=\hbar(k_x,k_y)$ is the crystal momentum from $K$-point.
The total Hamiltonian near $K$-point
for a field $\Psi_{\vec k}=(b_{1\vec k}, a_{2\vec k}, a_{1\vec k}, b_{2\vec k})^T$
can be written as
\begin{equation}
{\mathcal H}_{\vec k}=\hbar
\begin{pmatrix}
0       & v_3 k_- & v_0 k_+ & v_4 k_+ \\
v_3 k_+  &   0   & v_4 k_- & v_0 k_- \\
v_0 k_- & v_4 k_+ & 0 & -\gamma_1 \\
v_4 k_- & v_0 k_+ & -\gamma_1 & 0
\end{pmatrix},
\end{equation}
where
$k_{\pm}=k_x\pm i k_y$.
The effective Hamiltonian (${H}_{\rm eff}+H'_{\rm eff}$) 
on the low energy electronic structures
for a field $\Psi'_{\vec k} =(b_{1\vec k}, a_{2\vec k})^T$,
are described by
\begin{eqnarray}
{\mathcal H}_{\rm eff}&\simeq&
\hbar v_3 \begin{pmatrix}
0       & k_- \\
k_+  &   0
\end{pmatrix}
+\frac{\hbar^2 v^2_0}{\gamma_1}
\begin{pmatrix}
0       &  k^2_+ \\
k^2_-  &   0
\end{pmatrix},\\
{\mathcal H}'_{\rm eff}&\simeq&
\frac{2\hbar^2 v_0v_4}{\gamma_1}
\begin{pmatrix}
k_+ k_-       &  0 \\
0  & k_+ k_-
\end{pmatrix}\nonumber\\
& &+\frac{\hbar^2 v^2_4}{\gamma_1}
\begin{pmatrix}
0       &  k^2_+ \\
k^2_-  & 0
\end{pmatrix}.
\end{eqnarray}
Here we decompose the effective Hamiltonian into
${\mathcal H}_{\rm eff}$ and ${\mathcal H}'_{\rm eff}$ where the latter
is quite small compared to the former.
By using Pauli spin matrices,
$\tau_0=\left({1\atop0}{0\atop1}\right)$,
$\tau_x=\left({0\atop1}{1\atop0}\right)$, and
$\tau_y=\left({0\atop i}{-i\atop0}\right)$,
the above Eqs. (A.7) and (A.8) can be written in compact forms,
\begin{eqnarray}
{\mathcal H}_{\rm eff}&\simeq& \hbar v_3 \vec\tau\cdot\vec k
+\frac{\hbar^2 v^2_0}{\gamma_1}(\vec\tau^* \cdot\vec k)\tau_x (\vec\tau^*\cdot\vec k),\\
{\mathcal H}'_{\rm eff}&\simeq& \frac{2\hbar^2 v_0 v_4}{\gamma_1}|\vec k|^2 \tau_0 \nonumber\\
& &+\frac{\hbar^2 v^2_4}{\gamma_1}(\vec\tau^* \cdot \vec k)\tau_x (\vec\tau^*\cdot\vec k),
\end{eqnarray}
where ${\vec\tau}=(\tau_x,\tau_y)$ and $\vec\tau^*$ is its complex conjugate.
Hereafter, we will neglect the smallest
hopping parameter $\gamma_4$ (Eqs. (A.8) and (A.10)) and discuss its role later.

The effective Hamiltonian in Eq. (A.9) gives energy eigenvalues
\begin{equation}
E(\vec k)=\pm \hbar v_3 k\left|1 +\frac{k}{p_D}  e^{3i\phi_k}\right|,
\end{equation}
where $k=|\vec k|$, $\phi_k=\tan^{-1}\left({k_y}/{k_x}\right)$
and $p_D={\gamma_1 v_3}/({\hbar v_0^2})
=2/({3a_c}){\gamma_1 \gamma_3}/{\gamma_0^2}$.

When the layer 2 (bottom layer) slides against the layer 1 (top) 
along $\vec d_s=(d_x, d_y)$ ($|\vec d_s|\ll a_c$) [Figs. 1(a) and (b)],
the constant nnn inter-layer interaction ($\gamma_3$) now depends on the
carbon pair distance in layer 1 and 2 [Fig. 1(d)]
and nn inter-layer interaction ($\gamma_1$) decreases to $\gamma'_1$.
Since we focus on an extremely small sliding distance,
we assume variation of the nnn inter-layer interaction such
as $\gamma_3\rightarrow\gamma_3(\vec\delta_i)\simeq\gamma_3 e^{-\beta \vec\delta_i \cdot \vec d_s}
\simeq \gamma_3 (1-\beta\vec\delta_i\cdot\vec d_s)$
where $i=1,2,3$ and $\beta$ is a positive real constant.
When we expand Eq. (A.4) around $K$-point including sliding effect,
intra-layer interactions remain the same as before.
However, unlike the expansion procedure without sliding shown in Eq. (A.5),
the interlayer interaction with sliding can be expanded up to a leading order of $\vec k$ and $\vec d$ as 
\begin{eqnarray}
\xi_3 (\vec k+\vec K)&\simeq&
-\sum_j \gamma_3 (\vec\delta_j)e^{i(\vec k+\vec K)\cdot\vec\delta_j} \nonumber\\
&\simeq&-\sum_j \gamma_3 (1-\beta\vec\delta_i\cdot\vec d_s)
	(1+i\vec k\cdot \vec\delta_j)e^{i\vec K\cdot\vec\delta_j}\nonumber\\
&\simeq&\hbar v_3 (k_x +i k_y)+\hbar v_3 \beta (id_x -d_y).
\end{eqnarray}
Hence, the sliding vector $d_s$ plays a role of shifting the nnn intra-layer
Hamiltonian in momentum space like a constant vector potential.
With $\gamma_4$ neglected, the total Hamiltonian near $K$-point in Eq. (A.6)
now transforms to
\begin{equation}
{\mathcal H}_{\vec k}=\hbar
\begin{pmatrix}
0       & v_3 (k_- -\lambda_-)& v_0 k_+ &0  \\
v_3 (k_+-\lambda_+)  &   0   & 0 & v_0 k_- \\
v_0 k_- & 0 & 0 & -\gamma'_1 \\
0 & v_0 k_+ & -\gamma'_1 & 0
\end{pmatrix}
\end{equation}
where $\lambda_\pm=\beta(d_y\mp id_x)$.
Here we neglect an overall phase shift of $e^{i\vec d_s\cdot\vec k}$
because of $|\vec d_s|\ll a_c$.
The effective low energy Hamiltonian of Eq. (A.9) also
changes to
\begin{equation}
{\mathcal H}_{\rm eff}\simeq \hbar v_3 \vec\tau\cdot(\vec k -\vec \lambda)
+\frac{\hbar^2 v^2_0}{\gamma'_1}(\vec\tau^* \cdot\vec k)\tau_x (\vec\tau^*\cdot\vec k)
\end{equation}
where $\vec \lambda\equiv(\lambda_x,\lambda_y)=
\beta(d_y, -d_x)=\beta (\vec d_s \times \hat k_z)$ ($\hat k_z=\hat k_x \times \hat k_y$
and $(\hat k_x, \hat k_y)=\vec k/|\vec k|$).

Now, let us consider the smallest interlayer interaction $\gamma_4$ with sliding.
When sliding occurs, $\gamma_4$ shown in Fig. 1(e) 
becomes anisotropic and depends on the pair distances 
between relevant carbon atoms in the top and bottom layer.
We find that $\gamma_4\rightarrow \gamma_4(\vec\delta_i)\simeq \gamma_4 e^{\beta\vec\delta_i\cdot\vec d_s}
\simeq \gamma_4 (1+\beta \vec\delta_i\cdot\vec d_s)$ where $i=1,2,3$.
Here we assume the same coefficient $\beta$ of $\gamma_3 (\vec\delta_i)$ for calculation convenience
and note that it does not change the main conclusions of the paper.
Like Eq. (A.12) for $\gamma_3$, we can expand $\xi_4$ as
\begin{eqnarray}
\xi_4 (\vec k+\vec K)&\simeq&
-\sum_j \gamma_4 (\vec\delta_j)e^{i(\vec k+\vec K)\cdot\vec\delta_j}\nonumber\\
&\simeq&-\sum_j \gamma_4 (1+\beta\vec\delta_i\cdot\vec d_s)
	(1+i\vec k\cdot \vec\delta_j)e^{i\vec K\cdot\vec\delta_j}\nonumber\\
&\simeq&\hbar v_4 (k_x +i k_y)-\hbar v_4 \beta (id_x -d_y) \nonumber \\
	&=& \hbar v_4 k_+ +\hbar v_4 \lambda_+ .
\end{eqnarray}
With $\gamma_4$ included, 
now the total Hamiltonian near $K$-point in Eq. (A.13)
has an additional term, 
\begin{equation}
{\mathcal H}_{\vec k}=\hbar v_4
\begin{pmatrix}
0  & 0& 0 & k_+ + \lambda_+  \\
0  &  0& k_- + \lambda_-     & 0 \\
0 & k_+ + \lambda_+ & 0 &0 \\
k_- + \lambda_- & 0  & 0& 0 
\end{pmatrix}
\end{equation}

Then, Eq. (A.10) changes to
\begin{eqnarray}
{\mathcal H}'_{\rm eff}&\simeq& \frac{2\hbar^2 v_0 v_4}{\gamma_1}\vec k\cdot(\vec k+\vec\lambda) \tau_0 \nonumber \\
& + &\frac{\hbar^2 v^2_4}{\gamma_1}[\vec\tau^* \cdot (\vec k+\vec \lambda)]\tau_x [\vec\tau^*\cdot(\vec k+\vec\lambda)].
\end{eqnarray}


\begin{thebibliography}{50}
\expandafter\ifx\csname natexlab\endcsname\relax\def\natexlab#1{#1}\fi
\expandafter\ifx\csname bibnamefont\endcsname\relax
  \def\bibnamefont#1{#1}\fi
\expandafter\ifx\csname bibfnamefont\endcsname\relax
  \def\bibfnamefont#1{#1}\fi
\expandafter\ifx\csname citenamefont\endcsname\relax
  \def\citenamefont#1{#1}\fi
\expandafter\ifx\csname url\endcsname\relax
  \def\url#1{\texttt{#1}}\fi
\expandafter\ifx\csname urlprefix\endcsname\relax\def\urlprefix{URL }\fi
\providecommand{\bibinfo}[2]{#2}
\providecommand{\eprint}[2][]{\url{#2}}

\bibitem[{\citenamefont{Lifshitz}(1960)}]{lifshitz}
\bibinfo{author}{\bibfnamefont{I.~M.} \bibnamefont{Lifshitz}},
  \bibinfo{journal}{Sov. Phys. JETP} \textbf{\bibinfo{volume}{21}},
  \bibinfo{pages}{1130} (\bibinfo{year}{1960}).

\bibitem[{\citenamefont{Blanter et~al.}(1994)\citenamefont{Blanter, Kaganov,
  Pantsulaya, and Varlamov}}]{blanter}
\bibinfo{author}{\bibfnamefont{Y.~M.} \bibnamefont{Blanter}},
  \bibinfo{author}{\bibfnamefont{M.~I.} \bibnamefont{Kaganov}},
  \bibinfo{author}{\bibfnamefont{A.~V.} \bibnamefont{Pantsulaya}},
  \bibnamefont{and} \bibinfo{author}{\bibfnamefont{A.~A.}
  \bibnamefont{Varlamov}}, \bibinfo{journal}{Phys. Rep}
  \textbf{\bibinfo{volume}{245}}, \bibinfo{pages}{159} (\bibinfo{year}{1994}).

\bibitem[{\citenamefont{Varlamov et~al.}(1989)\citenamefont{Varlamov, Egorov,
  and Pantsulaya}}]{varlamov}
\bibinfo{author}{\bibfnamefont{A.~A.} \bibnamefont{Varlamov}},
  \bibinfo{author}{\bibfnamefont{V.~S.} \bibnamefont{Egorov}},
  \bibnamefont{and} \bibinfo{author}{\bibfnamefont{A.~V.}
  \bibnamefont{Pantsulaya}}, \bibinfo{journal}{Adv. Phys}
  \textbf{\bibinfo{volume}{38}}, \bibinfo{pages}{469} (\bibinfo{year}{1989}).

\bibitem[{\citenamefont{Armitage et~al.}(2010)\citenamefont{Armitage, Tediosi,
  L\'evy, Giann\'in\'i, Forro, and {van der Marel}}}]{armitage}
\bibinfo{author}{\bibfnamefont{N.~P.} \bibnamefont{Armitage}},
  \bibinfo{author}{\bibfnamefont{R.}~\bibnamefont{Tediosi}},
  \bibinfo{author}{\bibfnamefont{F.}~\bibnamefont{L\'evy}},
  \bibinfo{author}{\bibfnamefont{E.}~\bibnamefont{Giann\'in\'i}},
  \bibinfo{author}{\bibfnamefont{L.}~\bibnamefont{Forro}}, \bibnamefont{and}
  \bibinfo{author}{\bibfnamefont{D.}~\bibnamefont{{van der Marel}}},
  \bibinfo{journal}{Phys. Rev. Lett.} \textbf{\bibinfo{volume}{104}},
  \bibinfo{pages}{237401} (\bibinfo{year}{2010}).

\bibitem[{\citenamefont{Novoselov et~al.}(2006)\citenamefont{Novoselov, McCann,
  Morozov, Fal'ko, Katsnelson, Zeitler, Jiang, Shedin, and Geim}}]{novoselov}
\bibinfo{author}{\bibfnamefont{K.~S.} \bibnamefont{Novoselov}},
  \bibinfo{author}{\bibfnamefont{E.}~\bibnamefont{McCann}},
  \bibinfo{author}{\bibfnamefont{S.~V.} \bibnamefont{Morozov}},
  \bibinfo{author}{\bibfnamefont{V.~I.} \bibnamefont{Fal'ko}},
  \bibinfo{author}{\bibfnamefont{M.~I.} \bibnamefont{Katsnelson}},
  \bibinfo{author}{\bibfnamefont{U.}~\bibnamefont{Zeitler}},
  \bibinfo{author}{\bibfnamefont{D.}~\bibnamefont{Jiang}},
  \bibinfo{author}{\bibfnamefont{F.}~\bibnamefont{Shedin}}, \bibnamefont{and}
  \bibinfo{author}{\bibfnamefont{A.~K.} \bibnamefont{Geim}},
  \bibinfo{journal}{Nat. Phys.} \textbf{\bibinfo{volume}{2}},
  \bibinfo{pages}{177} (\bibinfo{year}{2006}).

\bibitem[{\citenamefont{Li et~al.}(2009)\citenamefont{Li, Luican, {Lopes dos
  Santos}, {Castro Neto}, Reina, Kong, and Andrei}}]{li}
\bibinfo{author}{\bibfnamefont{G.}~\bibnamefont{Li}},
  \bibinfo{author}{\bibfnamefont{A.}~\bibnamefont{Luican}},
  \bibinfo{author}{\bibfnamefont{J.~M.~B.} \bibnamefont{{Lopes dos Santos}}},
  \bibinfo{author}{\bibfnamefont{A.~H.} \bibnamefont{{Castro Neto}}},
  \bibinfo{author}{\bibfnamefont{A.}~\bibnamefont{Reina}},
  \bibinfo{author}{\bibfnamefont{J.}~\bibnamefont{Kong}}, \bibnamefont{and}
  \bibinfo{author}{\bibfnamefont{E.~Y.} \bibnamefont{Andrei}},
  \bibinfo{journal}{Nat. Phys.} \textbf{\bibinfo{volume}{6}},
  \bibinfo{pages}{109} (\bibinfo{year}{2009}).

\bibitem[{\citenamefont{Feldman et~al.}(2009)\citenamefont{Feldman, Martin, and
  Yacoby}}]{yacoby1}
\bibinfo{author}{\bibfnamefont{B.~E.} \bibnamefont{Feldman}},
  \bibinfo{author}{\bibfnamefont{J.}~\bibnamefont{Martin}}, \bibnamefont{and}
  \bibinfo{author}{\bibfnamefont{A.}~\bibnamefont{Yacoby}},
  \bibinfo{journal}{Nat. Phys.} \textbf{\bibinfo{volume}{5}},
  \bibinfo{pages}{889} (\bibinfo{year}{2009}).

\bibitem[{\citenamefont{Weitz et~al.}(2010)\citenamefont{Weitz, Allen, Feldman,
  Martin, and Yacoby}}]{yacoby2}
\bibinfo{author}{\bibfnamefont{R.~T.} \bibnamefont{Weitz}},
  \bibinfo{author}{\bibfnamefont{M.~T.} \bibnamefont{Allen}},
  \bibinfo{author}{\bibfnamefont{B.~E.} \bibnamefont{Feldman}},
  \bibinfo{author}{\bibfnamefont{J.}~\bibnamefont{Martin}}, \bibnamefont{and}
  \bibinfo{author}{\bibfnamefont{A.}~\bibnamefont{Yacoby}},
  \bibinfo{journal}{Science} \textbf{\bibinfo{volume}{330}},
  \bibinfo{pages}{812} (\bibinfo{year}{2010}).

\bibitem[{\citenamefont{Lemonik et~al.}(2010)\citenamefont{Lemonik, Aleiner,
  Toke, and Fal'ko}}]{lemonik}
\bibinfo{author}{\bibfnamefont{Y.}~\bibnamefont{Lemonik}},
  \bibinfo{author}{\bibfnamefont{I.~L.} \bibnamefont{Aleiner}},
  \bibinfo{author}{\bibfnamefont{C.}~\bibnamefont{Toke}}, \bibnamefont{and}
  \bibinfo{author}{\bibfnamefont{V.~I.} \bibnamefont{Fal'ko}},
  \bibinfo{journal}{Phys. Rev. B} \textbf{\bibinfo{volume}{82}},
  \bibinfo{pages}{201408} (\bibinfo{year}{2010}).

\bibitem[{\citenamefont{McCann and Fal'ko}(2006)}]{mccann}
\bibinfo{author}{\bibfnamefont{E.}~\bibnamefont{McCann}} \bibnamefont{and}
  \bibinfo{author}{\bibfnamefont{V.~I.} \bibnamefont{Fal'ko}},
  \bibinfo{journal}{Phys. Rev. Lett.} \textbf{\bibinfo{volume}{96}},
  \bibinfo{pages}{086805} (\bibinfo{year}{2006}).

\bibitem[{\citenamefont{Koshino and Ando}(2006)}]{ando}
\bibinfo{author}{\bibfnamefont{M.}~\bibnamefont{Koshino}} \bibnamefont{and}
  \bibinfo{author}{\bibfnamefont{T.}~\bibnamefont{Ando}},
  \bibinfo{journal}{Phys. Rev. B} \textbf{\bibinfo{volume}{73}},
  \bibinfo{pages}{245403} (\bibinfo{year}{2006}).

\bibitem[{\citenamefont{Nilsson et~al.}(2008)\citenamefont{Nilsson, {Castro
  Neto}, Guinea, and Peres}}]{nilsson}
\bibinfo{author}{\bibfnamefont{J.}~\bibnamefont{Nilsson}},
  \bibinfo{author}{\bibfnamefont{A.~H.} \bibnamefont{{Castro Neto}}},
  \bibinfo{author}{\bibfnamefont{F.}~\bibnamefont{Guinea}}, \bibnamefont{and}
  \bibinfo{author}{\bibfnamefont{N.~M.~R.} \bibnamefont{Peres}},
  \bibinfo{journal}{Phys. Rev. B} \textbf{\bibinfo{volume}{78}},
  \bibinfo{pages}{045405} (\bibinfo{year}{2008}).

\bibitem[{\citenamefont{Cserti et~al.}(2007)\citenamefont{Cserti, Csord{\'a}s,
  and D{\'a}vid}}]{cserti}
\bibinfo{author}{\bibfnamefont{J.}~\bibnamefont{Cserti}},
  \bibinfo{author}{\bibfnamefont{A.}~\bibnamefont{Csord{\'a}s}},
  \bibnamefont{and}
  \bibinfo{author}{\bibfnamefont{G.}~\bibnamefont{D{\'a}vid}},
  \bibinfo{journal}{Phys. Rev. Lett.} \textbf{\bibinfo{volume}{99}},
  \bibinfo{pages}{066802} (\bibinfo{year}{2007}).

\bibitem[{\citenamefont{Min et~al.}(2007)\citenamefont{Min, Sahu, Banerjee, and
  MacDonald}}]{min}
\bibinfo{author}{\bibfnamefont{H.}~\bibnamefont{Min}},
  \bibinfo{author}{\bibfnamefont{B.}~\bibnamefont{Sahu}},
  \bibinfo{author}{\bibfnamefont{S.~K.} \bibnamefont{Banerjee}},
  \bibnamefont{and} \bibinfo{author}{\bibfnamefont{A.~H.}
  \bibnamefont{MacDonald}}, \bibinfo{journal}{Phys. Rev. B}
  \textbf{\bibinfo{volume}{75}}, \bibinfo{pages}{155115}
  (\bibinfo{year}{2007}).

\bibitem[{\citenamefont{{Castro Neto} et~al.}(2009)\citenamefont{{Castro Neto},
  Guinea, Peres, Novoselov, and Geim}}]{neto}
\bibinfo{author}{\bibfnamefont{A.~H.} \bibnamefont{{Castro Neto}}},
  \bibinfo{author}{\bibfnamefont{F.}~\bibnamefont{Guinea}},
  \bibinfo{author}{\bibfnamefont{N.~M.~R.} \bibnamefont{Peres}},
  \bibinfo{author}{\bibfnamefont{K.~S.} \bibnamefont{Novoselov}},
  \bibnamefont{and} \bibinfo{author}{\bibfnamefont{A.~K.} \bibnamefont{Geim}},
  \bibinfo{journal}{Rev. Mod. Phys.} \textbf{\bibinfo{volume}{81}},
  \bibinfo{pages}{109} (\bibinfo{year}{2009}).

\bibitem[{\citenamefont{Castro et~al.}(2008)\citenamefont{Castro, Peres,
  Stauber, and Silva}}]{silva}
\bibinfo{author}{\bibfnamefont{E.~V.} \bibnamefont{Castro}},
  \bibinfo{author}{\bibfnamefont{N.~M.~R.} \bibnamefont{Peres}},
  \bibinfo{author}{\bibfnamefont{T.}~\bibnamefont{Stauber}}, \bibnamefont{and}
  \bibinfo{author}{\bibfnamefont{N.~A.~P.} \bibnamefont{Silva}},
  \bibinfo{journal}{Phys. Rev. Lett.} \textbf{\bibinfo{volume}{100}},
  \bibinfo{pages}{186803} (\bibinfo{year}{2008}).

\bibitem[{\citenamefont{Nandkishore and Levitov}(2010)}]{levitov}
\bibinfo{author}{\bibfnamefont{R.}~\bibnamefont{Nandkishore}} \bibnamefont{and}
  \bibinfo{author}{\bibfnamefont{L.}~\bibnamefont{Levitov}},
  \bibinfo{journal}{Phys. Rev. Lett.} \textbf{\bibinfo{volume}{104}},
  \bibinfo{pages}{156803} (\bibinfo{year}{2010}).

\bibitem[{\citenamefont{Zhang et~al.}(2010)\citenamefont{Zhang, Min, Polini,
  and MacDonald}}]{polini}
\bibinfo{author}{\bibfnamefont{F.}~\bibnamefont{Zhang}},
  \bibinfo{author}{\bibfnamefont{H.}~\bibnamefont{Min}},
  \bibinfo{author}{\bibfnamefont{M.}~\bibnamefont{Polini}}, \bibnamefont{and}
  \bibinfo{author}{\bibfnamefont{A.~H.} \bibnamefont{MacDonald}},
  \bibinfo{journal}{Phys. Rev. B} \textbf{\bibinfo{volume}{81}},
  \bibinfo{pages}{041402} (\bibinfo{year}{2010}).

\bibitem[{\citenamefont{Vafek and Yang}(2010)}]{vafek}
\bibinfo{author}{\bibfnamefont{O.}~\bibnamefont{Vafek}} \bibnamefont{and}
  \bibinfo{author}{\bibfnamefont{K.}~\bibnamefont{Yang}},
  \bibinfo{journal}{Phys. Rev. B} \textbf{\bibinfo{volume}{81}},
  \bibinfo{pages}{041401} (\bibinfo{year}{2010}).

\bibitem[{\citenamefont{{Lopes dos Santos} et~al.}(2007)\citenamefont{{Lopes
  dos Santos}, Peres, and {Castro Neto}}}]{lopes}
\bibinfo{author}{\bibfnamefont{J.~M.~B.} \bibnamefont{{Lopes dos Santos}}},
  \bibinfo{author}{\bibfnamefont{N.~M.~R.} \bibnamefont{Peres}},
  \bibnamefont{and} \bibinfo{author}{\bibfnamefont{A.~H.} \bibnamefont{{Castro
  Neto}}}, \bibinfo{journal}{Phys. Rev. Lett.} \textbf{\bibinfo{volume}{99}},
  \bibinfo{pages}{256802} (\bibinfo{year}{2007}).

\bibitem[{\citenamefont{Shallscross et~al.}(2008)\citenamefont{Shallscross,
  Sharma, and Pankratov}}]{shallcross}
\bibinfo{author}{\bibfnamefont{S.}~\bibnamefont{Shallscross}},
  \bibinfo{author}{\bibfnamefont{S.}~\bibnamefont{Sharma}}, \bibnamefont{and}
  \bibinfo{author}{\bibfnamefont{O.~A.} \bibnamefont{Pankratov}},
  \bibinfo{journal}{Phys. Rev. Lett.} \textbf{\bibinfo{volume}{101}},
  \bibinfo{pages}{056803} (\bibinfo{year}{2008}).

\bibitem[{\citenamefont{Hass et~al.}(2008)\citenamefont{Hass, Varchon,
  Mill'{a}-Otoya, Sprinkle, Sharma, de~Heer, Berger, First, Magaud, and
  Conrad}}]{hass}
\bibinfo{author}{\bibfnamefont{J.}~\bibnamefont{Hass}},
  \bibinfo{author}{\bibfnamefont{F.}~\bibnamefont{Varchon}},
  \bibinfo{author}{\bibfnamefont{J.~E.} \bibnamefont{Mill'{a}-Otoya}},
  \bibinfo{author}{\bibfnamefont{M.}~\bibnamefont{Sprinkle}},
  \bibinfo{author}{\bibfnamefont{N.}~\bibnamefont{Sharma}},
  \bibinfo{author}{\bibfnamefont{W.~A.} \bibnamefont{de~Heer}},
  \bibinfo{author}{\bibfnamefont{C.}~\bibnamefont{Berger}},
  \bibinfo{author}{\bibfnamefont{P.~N.} \bibnamefont{First}},
  \bibinfo{author}{\bibfnamefont{L.}~\bibnamefont{Magaud}}, \bibnamefont{and}
  \bibinfo{author}{\bibfnamefont{E.~H.} \bibnamefont{Conrad}},
  \bibinfo{journal}{Phys. Rev. Lett.} \textbf{\bibinfo{volume}{100}},
  \bibinfo{pages}{125504} (\bibinfo{year}{2008}).

\bibitem[{\citenamefont{Giannozzi et~al.}(2009)\citenamefont{Giannozzi, Baroni,
  Bonini, Calandra, Car, Cavazzoni, Ceresoli, Chiarotti, Cococcioni, Dabo
  et~al.}}]{espresso}
\bibinfo{author}{\bibfnamefont{P.}~\bibnamefont{Giannozzi}},
  \bibinfo{author}{\bibfnamefont{S.}~\bibnamefont{Baroni}},
  \bibinfo{author}{\bibfnamefont{N.}~\bibnamefont{Bonini}},
  \bibinfo{author}{\bibfnamefont{M.}~\bibnamefont{Calandra}},
  \bibinfo{author}{\bibfnamefont{R.}~\bibnamefont{Car}},
  \bibinfo{author}{\bibfnamefont{C.}~\bibnamefont{Cavazzoni}},
  \bibinfo{author}{\bibfnamefont{D.}~\bibnamefont{Ceresoli}},
  \bibinfo{author}{\bibfnamefont{G.~L.} \bibnamefont{Chiarotti}},
  \bibinfo{author}{\bibfnamefont{M.}~\bibnamefont{Cococcioni}},
  \bibinfo{author}{\bibfnamefont{I.}~\bibnamefont{Dabo}}, 
  A. Dal Corso, S. de Cironcoli, S. Fabris, G. Fratesi, R. Gebauer, 
  U. Gerstman, C. Gougoussis, A.Kokalj, M. Lazzeri, L. Martin-Samos,
  N. Marzari, F. Mauri, R. Mazzarello, S. Paolini, A. Pasquarello, 
  L. Paulatto, C. Sbraccia, S. Scandolo, G. Sclauzero, A. P. Seitsonen, 
  A. Smogunov, P. Umari,and R. M. Wentzcovitch,
  \bibinfo{journal}{J. Phys.:Condens. Matter} \textbf{\bibinfo{volume}{21}},
  \bibinfo{pages}{395502} (\bibinfo{year}{2009}).

\bibitem[{\citenamefont{Perdew et~al.}(1996)\citenamefont{Perdew, Burke, and
  Ernzerhof}}]{pbe}
\bibinfo{author}{\bibfnamefont{J.~P.} \bibnamefont{Perdew}},
  \bibinfo{author}{\bibfnamefont{K.}~\bibnamefont{Burke}}, \bibnamefont{and}
  \bibinfo{author}{\bibfnamefont{M.}~\bibnamefont{Ernzerhof}},
  \bibinfo{journal}{Phys. Rev. Lett.} \textbf{\bibinfo{volume}{77}},
  \bibinfo{pages}{3865} (\bibinfo{year}{1996}).

\bibitem[{\citenamefont{Vanderbilt}(1990)}]{vanderbilt}
\bibinfo{author}{\bibfnamefont{D.}~\bibnamefont{Vanderbilt}},
  \bibinfo{journal}{Phys. Rev. B} \textbf{\bibinfo{volume}{41}},
  \bibinfo{pages}{7892} (\bibinfo{year}{1990}).

\bibitem[{\citenamefont{Marzari et~al.}(1999)\citenamefont{Marzari, Vanderbilt,
  {De Vita}, and Payne}}]{marzari}
\bibinfo{author}{\bibfnamefont{N.}~\bibnamefont{Marzari}},
  \bibinfo{author}{\bibfnamefont{D.}~\bibnamefont{Vanderbilt}},
  \bibinfo{author}{\bibfnamefont{A.}~\bibnamefont{{De Vita}}},
  \bibnamefont{and} \bibinfo{author}{\bibfnamefont{M.~C.} \bibnamefont{Payne}},
  \bibinfo{journal}{Phys. Rev. Lett.} \textbf{\bibinfo{volume}{82}},
  \bibinfo{pages}{3296} (\bibinfo{year}{1999}).

\bibitem[{\citenamefont{Kresse and Hafner}(1994)}]{kresse}
\bibinfo{author}{\bibfnamefont{G.}~\bibnamefont{Kresse}} \bibnamefont{and}
  \bibinfo{author}{\bibfnamefont{J.}~\bibnamefont{Hafner}},
  \bibinfo{journal}{Phys. Rev. B} \textbf{\bibinfo{volume}{49}},
  \bibinfo{pages}{14251} (\bibinfo{year}{1994}).

\bibitem[{\citenamefont{Grimme}(2006)}]{grimme}
\bibinfo{author}{\bibfnamefont{S.}~\bibnamefont{Grimme}}, \bibinfo{journal}{J.
  Comput. Chem.} \textbf{\bibinfo{volume}{27}}, \bibinfo{pages}{1787}
  (\bibinfo{year}{2006}).

\bibitem[{\citenamefont{Barone et~al.}(2008)\citenamefont{Barone, Casarin,
  Forrer, Pavone, Sambi, and Vittadini}}]{barone}
\bibinfo{author}{\bibfnamefont{V.}~\bibnamefont{Barone}},
  \bibinfo{author}{\bibfnamefont{M.}~\bibnamefont{Casarin}},
  \bibinfo{author}{\bibfnamefont{D.}~\bibnamefont{Forrer}},
  \bibinfo{author}{\bibfnamefont{M.}~\bibnamefont{Pavone}},
  \bibinfo{author}{\bibfnamefont{M.}~\bibnamefont{Sambi}}, \bibnamefont{and}
  \bibinfo{author}{\bibfnamefont{A.}~\bibnamefont{Vittadini}},
  \bibinfo{journal}{J. Comput. Chem.} \textbf{\bibinfo{volume}{30}},
  \bibinfo{pages}{934} (\bibinfo{year}{2008}).

\bibitem[{\citenamefont{Kolmogorov and Crespi}(2000)}]{crespi}
\bibinfo{author}{\bibfnamefont{A.~N.} \bibnamefont{Kolmogorov}}
  \bibnamefont{and} \bibinfo{author}{\bibfnamefont{V.~H.}
  \bibnamefont{Crespi}}, \bibinfo{journal}{Phys. Rev. Lett.}
  \textbf{\bibinfo{volume}{85}}, \bibinfo{pages}{4727} (\bibinfo{year}{2000}).

\bibitem[{\citenamefont{Wu and Yang}(2002)}]{yang}
\bibinfo{author}{\bibfnamefont{Q.}~\bibnamefont{Wu}} \bibnamefont{and}
  \bibinfo{author}{\bibfnamefont{W.}~\bibnamefont{Yang}}, \bibinfo{journal}{J.
  Chem. Phys.} \textbf{\bibinfo{volume}{116}}, \bibinfo{pages}{515}
  (\bibinfo{year}{2002}).

\bibitem[{\citenamefont{Gould et~al.}(2008)\citenamefont{Gould, Simpkins, and
  Dobson}}]{dobson}
\bibinfo{author}{\bibfnamefont{T.}~\bibnamefont{Gould}},
  \bibinfo{author}{\bibfnamefont{K.}~\bibnamefont{Simpkins}}, \bibnamefont{and}
  \bibinfo{author}{\bibfnamefont{J.~F.} \bibnamefont{Dobson}},
  \bibinfo{journal}{Phys. Rev. B} \textbf{\bibinfo{volume}{77}},
  \bibinfo{pages}{165134} (\bibinfo{year}{2008}).

\bibitem[{\citenamefont{Spanu et~al.}(2009)\citenamefont{Spanu, Sorella, and
  Galli}}]{galli}
\bibinfo{author}{\bibfnamefont{L.}~\bibnamefont{Spanu}},
  \bibinfo{author}{\bibfnamefont{S.}~\bibnamefont{Sorella}}, \bibnamefont{and}
  \bibinfo{author}{\bibfnamefont{G.}~\bibnamefont{Galli}},
  \bibinfo{journal}{Phys. Rev. Lett.} \textbf{\bibinfo{volume}{103}},
  \bibinfo{pages}{196401} (\bibinfo{year}{2009}).

\bibitem[{\citenamefont{Grimme et~al.}(2007)\citenamefont{Grimme,
  M\"uck-Lichtenfeld, and Antony}}]{antony}
\bibinfo{author}{\bibfnamefont{S.}~\bibnamefont{Grimme}},
  \bibinfo{author}{\bibfnamefont{G.}~\bibnamefont{M\"uck-Lichtenfeld}},
  \bibnamefont{and} \bibinfo{author}{\bibfnamefont{J.}~\bibnamefont{Antony}},
  \bibinfo{journal}{J. Phys. Chem. C} \textbf{\bibinfo{volume}{111}},
  \bibinfo{pages}{11199} (\bibinfo{year}{2007}).

\bibitem[{\citenamefont{Chakarova-K\"ack
  et~al.}(2006)\citenamefont{Chakarova-K\"ack, Schr\"oder, Lundqvist, and
  Langreth}}]{langreth}
\bibinfo{author}{\bibfnamefont{S.~D.} \bibnamefont{Chakarova-K\"ack}},
  \bibinfo{author}{\bibfnamefont{E.}~\bibnamefont{Schr\"oder}},
  \bibinfo{author}{\bibfnamefont{B.~I.} \bibnamefont{Lundqvist}},
  \bibnamefont{and} \bibinfo{author}{\bibfnamefont{D.~C.}
  \bibnamefont{Langreth}}, \bibinfo{journal}{Phys. Rev. Lett.}
  \textbf{\bibinfo{volume}{96}}, \bibinfo{pages}{146107}
  (\bibinfo{year}{2006}).

\bibitem[{\citenamefont{Kuzmenko et~al.}(2009)\citenamefont{Kuzmenko, Crassee,
  {van der Marel}, Blake, and Novoselov}}]{kuzmenko}
\bibinfo{author}{\bibfnamefont{A.~B.} \bibnamefont{Kuzmenko}},
  \bibinfo{author}{\bibfnamefont{I.}~\bibnamefont{Crassee}},
  \bibinfo{author}{\bibfnamefont{D.}~\bibnamefont{{van der Marel}}},
  \bibinfo{author}{\bibfnamefont{P.}~\bibnamefont{Blake}}, \bibnamefont{and}
  \bibinfo{author}{\bibfnamefont{K.~S.} \bibnamefont{Novoselov}},
  \bibinfo{journal}{Phys. Rev. B} \textbf{\bibinfo{volume}{80}},
  \bibinfo{pages}{165406} (\bibinfo{year}{2009}).

\bibitem[{\citenamefont{Pereira et~al.}(2009)\citenamefont{Pereira, Neto, and
  Peres}}]{pereira}
\bibinfo{author}{\bibfnamefont{V.~M.} \bibnamefont{Pereira}},
  \bibinfo{author}{\bibfnamefont{A.~H.} \bibnamefont{Castro Neto}},
  \bibnamefont{and} \bibinfo{author}{\bibfnamefont{N.~M.~R.}
  \bibnamefont{Peres}}, \bibinfo{journal}{Phys. Rev. B}
  \textbf{\bibinfo{volume}{80}}, \bibinfo{pages}{045401}
  (\bibinfo{year}{2009}).

\bibitem[{\citenamefont{Choi et~al.}(2010{\natexlab{a}})\citenamefont{Choi,
  Jhi, and Son}}]{choi}
\bibinfo{author}{\bibfnamefont{S.-M.} \bibnamefont{Choi}},
  \bibinfo{author}{\bibfnamefont{S.-H.} \bibnamefont{Jhi}}, \bibnamefont{and}
  \bibinfo{author}{\bibfnamefont{Y.-W.} \bibnamefont{Son}},
  \bibinfo{journal}{Phys. Rev. B} \textbf{\bibinfo{volume}{81}},
  \bibinfo{pages}{081407} (\bibinfo{year}{2010}{\natexlab{a}}).

\bibitem[{\citenamefont{Guinea et~al.}(2010)\citenamefont{Guinea, Katsnelson,
  and Geim}}]{guinea2}
\bibinfo{author}{\bibfnamefont{F.}~\bibnamefont{Guinea}},
  \bibinfo{author}{\bibfnamefont{M.~I.} \bibnamefont{Katsnelson}},
  \bibnamefont{and} \bibinfo{author}{\bibfnamefont{A.~K.} \bibnamefont{Geim}},
  \bibinfo{journal}{Nat. Phys.} \textbf{\bibinfo{volume}{6}},
  \bibinfo{pages}{30} (\bibinfo{year}{2010}).

\bibitem[{\citenamefont{Choi et~al.}(2010{\natexlab{b}})\citenamefont{Choi,
  Jhi, and Son}}]{choi2}
\bibinfo{author}{\bibfnamefont{S.-M.} \bibnamefont{Choi}},
  \bibinfo{author}{\bibfnamefont{S.-H.} \bibnamefont{Jhi}}, \bibnamefont{and}
  \bibinfo{author}{\bibfnamefont{Y.-W.} \bibnamefont{Son}},
  \bibinfo{journal}{Nano Lett.} \textbf{\bibinfo{volume}{10}},
  \bibinfo{pages}{3486} (\bibinfo{year}{2010}{\natexlab{b}}).

\bibitem[{\citenamefont{Song et~al.}(2010)\citenamefont{Song, Otte, Kuk, Hu,
  Torrance, First, de~Heer, Min, Adam, Stiles et~al.}}]{song}
\bibinfo{author}{\bibfnamefont{Y.~J.} \bibnamefont{Song}},
  \bibinfo{author}{\bibfnamefont{A.~F.} \bibnamefont{Otte}},
  \bibinfo{author}{\bibfnamefont{Y.}~\bibnamefont{Kuk}},
  \bibinfo{author}{\bibfnamefont{Y.}~\bibnamefont{Hu}},
  \bibinfo{author}{\bibfnamefont{D.~B.} \bibnamefont{Torrance}},
  \bibinfo{author}{\bibfnamefont{P.~N.} \bibnamefont{First}},
  \bibinfo{author}{\bibfnamefont{W.~A.} \bibnamefont{de~Heer}},
  \bibinfo{author}{\bibfnamefont{H.}~\bibnamefont{Min}},
  \bibinfo{author}{\bibfnamefont{S.}~\bibnamefont{Adam}},
  \bibinfo{author}{\bibfnamefont{M.~D.} \bibnamefont{Stiles}},
 A. H. MacDonald and J. A. Stroscio,
   \bibinfo{journal}{Nature}
  \textbf{\bibinfo{volume}{467}}, \bibinfo{pages}{185} (\bibinfo{year}{2010}).

\bibitem[{\citenamefont{Montambaux et~al.}(2009)\citenamefont{Montambaux,
  Pi\'echon, Fuchs, and Goerbig}}]{goerbig}
\bibinfo{author}{\bibfnamefont{G.}~\bibnamefont{Montambaux}},
  \bibinfo{author}{\bibfnamefont{F.}~\bibnamefont{Pi\'echon}},
  \bibinfo{author}{\bibfnamefont{J.-N.} \bibnamefont{Fuchs}}, \bibnamefont{and}
  \bibinfo{author}{\bibfnamefont{M.~O.} \bibnamefont{Goerbig}},
  \bibinfo{journal}{Phys. Rev. B} \textbf{\bibinfo{volume}{80}},
  \bibinfo{pages}{153412} (\bibinfo{year}{2009}).

\bibitem[{\citenamefont{Wunsch et~al.}(2008)\citenamefont{Wunsch, Guinea, and
  Sols}}]{wunsch}
\bibinfo{author}{\bibfnamefont{B.}~\bibnamefont{Wunsch}},
  \bibinfo{author}{\bibfnamefont{F.}~\bibnamefont{Guinea}}, \bibnamefont{and}
  \bibinfo{author}{\bibfnamefont{F.}~\bibnamefont{Sols}},
  \bibinfo{journal}{New. J. Phys.} \textbf{\bibinfo{volume}{10}},
  \bibinfo{pages}{103027} (\bibinfo{year}{2008}).

\bibitem[{\citenamefont{Mikitik and Sharlai}(1999)}]{mikitik}
\bibinfo{author}{\bibfnamefont{G.~P.} \bibnamefont{Mikitik}} \bibnamefont{and}
  \bibinfo{author}{\bibfnamefont{Y.~V.} \bibnamefont{Sharlai}},
  \bibinfo{journal}{Phys. Rev. Lett.} \textbf{\bibinfo{volume}{82}},
  \bibinfo{pages}{2147} (\bibinfo{year}{1999}).

\bibitem[{\citenamefont{Dietal et~al.}(2008)\citenamefont{Dietal, Pi\'echon,
  and Montambaux}}]{dietl}
\bibinfo{author}{\bibfnamefont{P.}~\bibnamefont{Dietal}},
  \bibinfo{author}{\bibfnamefont{F.}~\bibnamefont{Pi\'echon}},
  \bibnamefont{and}
  \bibinfo{author}{\bibfnamefont{G.}~\bibnamefont{Montambaux}},
  \bibinfo{journal}{Phys. Rev. Lett.} \textbf{\bibinfo{volume}{100}},
  \bibinfo{pages}{236405} (\bibinfo{year}{2008}).

\bibitem[{\citenamefont{Bermudez et~al.}(2010)\citenamefont{Bermudez, Goldman,
  Kubasiak, Lewenstein, and Martin-Delgado}}]{bermudez}
\bibinfo{author}{\bibfnamefont{A.}~\bibnamefont{Bermudez}},
  \bibinfo{author}{\bibfnamefont{N.}~\bibnamefont{Goldman}},
  \bibinfo{author}{\bibfnamefont{A.}~\bibnamefont{Kubasiak}},
  \bibinfo{author}{\bibfnamefont{M.}~\bibnamefont{Lewenstein}},
  \bibnamefont{and} \bibinfo{author}{\bibfnamefont{M.~A.}
  \bibnamefont{Martin-Delgado}}, \bibinfo{journal}{New. J. Phys.}
  \textbf{\bibinfo{volume}{12}}, \bibinfo{pages}{033041}
  (\bibinfo{year}{2010}).

\bibitem[{\citenamefont{Hosotani}(1983)}]{hosotani}
\bibinfo{author}{\bibfnamefont{Y.}~\bibnamefont{Hosotani}},
  \bibinfo{journal}{Phys. Lett.} \textbf{\bibinfo{volume}{129B}},
  \bibinfo{pages}{193} (\bibinfo{year}{1983}).

\bibitem[{\citenamefont{Mucha-Kruczynski
  et~al.}(2011)\citenamefont{Mucha-Kruczynski, Aleiner, and Fal'ko}}]{marcin}
\bibinfo{author}{\bibfnamefont{M.}~\bibnamefont{Mucha-Kruczynski}},
  \bibinfo{author}{\bibfnamefont{I.~L.} \bibnamefont{Aleiner}},
  \bibnamefont{and} \bibinfo{author}{\bibfnamefont{V.~I.}
  \bibnamefont{Fal'ko}}, \bibinfo{journal}{Phys. Rev. B}
  \textbf{\bibinfo{volume}{84}}, \bibinfo{pages}{041404}
  (\bibinfo{year}{2011}).

\bibitem[{\citenamefont{de~Gail et~al.}(2011)\citenamefont{de~Gail, Goerbig,
  Guinea, Montambaux, and Neto}}]{gail}
\bibinfo{author}{\bibfnamefont{R.}~\bibnamefont{de~Gail}},
  \bibinfo{author}{\bibfnamefont{M.~O.} \bibnamefont{Goerbig}},
  \bibinfo{author}{\bibfnamefont{F.}~\bibnamefont{Guinea}},
  \bibinfo{author}{\bibfnamefont{G.}~\bibnamefont{Montambaux}},
  \bibnamefont{and} \bibinfo{author}{\bibfnamefont{A.~H.} \bibnamefont{Castro Neto}}, \bibinfo{journal}{Phys. Rev. B}
  \textbf{\bibinfo{volume}{84}}, \bibinfo{pages}{045436}
  (\bibinfo{year}{2011}).

\bibitem[{\citenamefont{S.Mayorov et~al.}(2011)\citenamefont{S.Mayorov, Elias,
  Mucha-Kruczynski, Gorbachev, Tudorovskiy, Zhukov, Katsnelson, Fal'ko, Geim,
  and Novoselov}}]{mayorov}
\bibinfo{author}{\bibfnamefont{A.}~\bibnamefont{S. Mayorov}},
  \bibinfo{author}{\bibfnamefont{D.~C.} \bibnamefont{Elias}},
  \bibinfo{author}{\bibfnamefont{M.}~\bibnamefont{Mucha-Kruczynski}},
  \bibinfo{author}{\bibfnamefont{R.~V.} \bibnamefont{Gorbachev}},
  \bibinfo{author}{\bibfnamefont{T.}~\bibnamefont{Tudorovskiy}},
  \bibinfo{author}{\bibfnamefont{A.}~\bibnamefont{Zhukov}},
  \bibinfo{author}{\bibfnamefont{S.~V. Morozov,} \bibnamefont{M.~I. Katsnelson}},
  \bibinfo{author}{\bibfnamefont{V.~I.} \bibnamefont{Fal'ko}},
  \bibinfo{author}{\bibfnamefont{A.~K.} \bibnamefont{Geim}}, \bibnamefont{and}
  \bibinfo{author}{\bibfnamefont{K.~S.} \bibnamefont{Novoselov}},
  \bibinfo{journal}{Science} \textbf{\bibinfo{volume}{333}},
  \bibinfo{pages}{860} (\bibinfo{year}{2011}).

\end{thebibliography}
\end{document}